\documentclass[twocolumn,showpacs,preprintnumbers,amsmath,amssymb]{revtex4}

\usepackage{graphicx}
\usepackage[usenames,dvipsnames]{xcolor} 
\usepackage[latin1]{inputenc}
\usepackage[T1]{fontenc}
\usepackage{latexsym}
\usepackage{amsfonts}
\usepackage{amssymb}
\usepackage{amsmath}
\usepackage{color}
\usepackage{psfrag}
\usepackage{rotating}
\usepackage{pgf,pgfsys,pgffor}
\usepackage{pgfplots}
\usepackage{pgfplotstable}
\usepackage[caption=false]{subfig} 
\usepackage{tikz}
\usetikzlibrary{arrows}
\usepackage{steinmetz}
\usepackage{trfsigns}
\usetikzlibrary{decorations.shapes}
\usepackage{hyperref}

\graphicspath{{./figureColor/}} 

\newcommand{\tr}[2][]{\textrm{Tr}_{#1} \left[ {#2} \right]}

\newcommand{\ket}[1]{\vert {#1} \rangle}
\newcommand{\pure}[1]{\vert {#1} \rangle \langle {#1} \vert}
\newcommand{\inn}[2]{\langle {#1} \vert {#2} \rangle}
\newcommand{\san}[3]{\langle {#1} \vert {#2} \vert {#3} \rangle}
\renewcommand{\e}{\textrm{e}}			
\newcommand{\imaginary}{\im}

\newcommand{\beq}{\begin{equation}}
\newcommand{\eeq}{\end{equation}}
\renewcommand{\d}{\textrm{d}} 

\newtheorem{cor}{Corollary}
\newtheorem{Lemma}{Lemma}

\newcommand{\proof}{\noindent {\bf Proof. }}
\newcommand{\qed}{\hfill $\Box$ \vskip 2ex}

\newcommand{\barg}{\bar{\gamma}} 
\newcommand{\coherentTensorProduct}{\gamma} 
\newcommand{\g}{\xi}	
\newcommand{\m}{\mu}	
\newcommand{\M}{M}	
\newcommand{\Prob}[1]{\textrm{P} \left[ #1 \right]} 
\newcommand{\p}{p} 
\newcommand{\pProj}{\hat{p}} 
\newcommand{\q}{q} 
\newcommand{\x}{x} 
\newcommand{\y}{y} 
\newcommand{\z}{z} 
\newcommand{\barz}{\bar{\z}} 
\newcommand{\barp}{\bar{\p}} 
\newcommand{\X}{\textrm{x}} \newcommand{\Y}{\textrm{y}} 
\newcommand{\inner}{\chi} 
\renewcommand{\o}{0} 
\renewcommand{\l}{1}
\newcommand{\newO}{\hat{\o}}
\newcommand{\newL}{\hat{\l}}
\newcommand{\Pc}{P_c}
\newcommand{\Pe}{P_e}
\newcommand{\povm}{\textrm{P}} 
\newcommand{\zM}{\textrm{z}} 
\newcommand{\f}{h} 
\newcommand{\mx}{m} 
\newcommand{\ang}{\phi}
\newcommand{\angDP}{\pi}
\newcommand{\nextS}{\textrm{children}}
\newcommand{\pro}{\wp} 
\newcommand{\proj}{\Pi} 

\newcommand{\s}{s}		
\renewcommand{\S}{S} 		
\renewcommand{\u}{u}		
\newcommand{\U}{U} 		
\newcommand{\w}{w}		
\newcommand{\W}{W} 		
\newcommand{\E}[2][]{\operatorname*{\textrm{E}}_{#1}\left[ #2 \right]} 
\newcommand{\J}{J} 		
\newcommand{\Jc}{\tilde{J}} 		
\newcommand{\fTable}{\Jc}

\begin{document}

\title{Adaptive Discrimination Scheme for Quantum Pulse Position Modulation Signals}


\author{Nicola Dalla Pozza}
  \email{nicola.dallapozza@dei.unipd.it}
  \affiliation{Department of Information Engineering (DEI), University of Padova}%

\author{Nicola Laurenti}%
  \email{nil@dei.unipd.it}
  \affiliation{Department of Information Engineering (DEI), University of Padova}%
  
\date{\today}
             
\begin{abstract}
In the communication scenario, we consider the problem of the discrimination between the signals of the Quantum Pulse Position Modulation. 
We propose a receiver scheme that employs repeated local measurements in distinct temporal slots within the symbol time interval, adaptively chosen on the basis of the outcomes in the previous slots. By employing Dynamic Programming to optimize each measurement, we approach the theoretical performance limit much closer than existing receiver schemes.
\end{abstract}


\maketitle

\section{\label{introduction}Introduction}

The increasing demand for higher data rates in free space communications is driving system designers to consider optical frequencies, rather then radio carriers, for the satellite-to-satellite and satellite-to-earth links. In this switch of design paradigm, quantum optics plays an important role for the correct description of the physical phenomena involved in the transmission, propagation and detection stages of the communication. 

Quantum optics supports the use of coherent states for the transmission of information over free space links with a long list of studies and results in both communication and information theory \cite{Weedbrook2012}.
For example, in \cite{Giovannetti2004} it is proven that with a random coding over coherent states it is possible to achieve the capacity of a lossy bosonic link, that is the quantum model for the free space channel. Furthermore, binary coherent state discrimination has been theoretically solved \cite{Dolinar1973} and experimentally tested \cite{Cook2007}, reaching the ultimate quantum limit (Helstrom bound \cite{Helstrom1976}) for the performance in terms of error probability.


Different solutions have been proposed to encode information into coherent states, most notably amplitude modulation, phase modulation and pulse position modulation. The choice depends upon several factors to be taken into account in the design of the communication system, including the channel model, the target performance, and the complexity of the system at the transmitter and at the receiver side. 

\begin{figure}
\centering
	\includegraphics{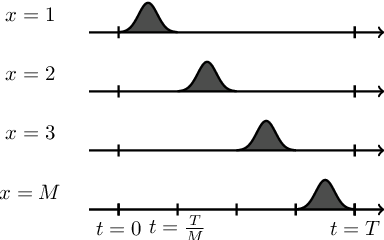}
  \caption{\label{fig:pulse}Pulse Position Modulation, association between symbols $\x$ and the position of the pulse in the symbol time interval, with alphabet cardinality $\M=4$.}
\end{figure}

Pulse Position Modulation (PPM) encodes the information to be transmitted in the temporal position of a pulse within the symbol time length. 
For example, Figure~\ref{fig:pulse} depicts the mapping from the symbols $\x~=~\{1,2,3,4\}$ of a 4-PPM to the transmitted signal. The fact that the implementation of a PPM transmitter only requires an intensity modulator, and its efficiency in terms of average energy
make this modulation a suitable candidate for communications from satellites and spacecrafts.

The classical approach of the receiver for the discrimination of PPM signals is to detect the field intensity in the symbol time interval, and  to estimate the transmitted symbol according to the slot where the maximum of energy has been measured. Dolinar \cite{Dolinar1983} proposed an adaptive receiver scheme that exploits the possibility of nulling the PPM signal depending on the result of an intensity measurement.
Recently, an improved version of this receiver scheme has been introduced \cite{Guha2011}, and has received a lot of attention due to the possibility to outperform the standard quantum limit given by the photon counting and approach the theoretical limit predicted by quantum discrimination theory (Helstrom bound). This solution has also been experimentally tested and tuned to face experimental non idealities \cite{Chen2012}.

In this paper we briefly review these receiver schemes and propose a similar feedback structure consisting of a sequence of $\M$ local measurements optimized by means of dynamic programming. This optimization allows to approach the Helstrom bound more closely than existing receiver schemes. 

The paper is organized as follows. 
In Section \ref{communication}, we describe the communication setup and formally state the problem.
In Section \ref{structure} we review the existing structures for PPM receivers, and propose our adaptive scheme that will be optimized later.
Section \ref{review} briefly reviews some results in dynamic programming, which are then employed in Section \ref{optimization} to derive the procedure for the optimization of the receiver scheme.
In Section \ref{results} we show the performance of the receiver scheme and describe some numerical issues in the optimization procedure.
Section \ref{conclusions} summarizes the paper contribution and draws conclusions.

%
%
%

\section{\label{communication}Communication Setup}

\begin{figure}[h]
\centering
\includegraphics{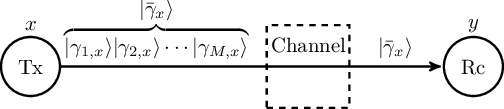}
\caption{\label{fig:scenario}Communication system setup, with ideal channel. With this assumption, the transmitted quantum states are received unaltered.}
\end{figure}

We consider the communication system summarized in Figure \ref{fig:scenario}. A transmitter encodes its message in a sequence of symbols $\x \in \{1,\ldots, \M\}$, described as a random variable with distribution $\{\p_1, \ldots , \p_{\M}\}$. We assume an equal a priori distribution for the symbols $\x$, i.e. $\p_{\x}=\frac{1}{\M} \ \forall \x$. 

The transmitter maps each symbol of the sequence to a quantum state $\ket{\barg_\x}$ taken from the set $\{\ket{\barg_1}, \ket{\barg_2}, \dots, \ket{\barg_{\M}} \}$ and sends it through the channel. We assume the channel to be ideal, such that the quantum states at the channel output are exactly the ones that have been transmitted. 

The receiver measures the output of the quantum channel and estimates which symbol $\x$ has been encoded. We denote with $\y$ the random variable associated with the symbol estimation. The figure of merit to evaluate the performance of the receiver  scheme is the probability of correct decision, or equivalently, the error probability, defined from transmitted and estimated symbol as
\begin{align}
\Pc 	= \sum_{i=1}^{\M} \Prob{\y=i,\ \x=i} &= \frac{1}{\M} \sum_{i=1}^{\M} \Prob{\y=i| \x=i}, \label{def:Pc} \\
\Pe & = 1-\Pc. \label{def:Pe}
\end{align}

Pulse Position Modulation defines a particular structure for the quantum states $\ket{\barg_\x}$. As we can see from Figure \ref{fig:pulse}, the symbol time interval can be virtually divided in $\M$ temporal slots. A pulse in the $i$-th slot is associated to the symbol $\x=i$. Quantum optics describes this signal as a sequence of coherent states in a tensor product, one in each slot, with all ground states $\ket{0}$ except the coherent state $\ket{\alpha}$ in the $i$-th position, $\alpha \neq 0$. 
\beq
\ket{\barg_\x} = 
\ket{\coherentTensorProduct_{1,\x}}\ket{\coherentTensorProduct_{2,\x}} \ldots \ket{\coherentTensorProduct_{\M,\x}}, 
\quad \coherentTensorProduct_{i,\x} = 
\begin{cases}
\alpha & i=\x \\
0 & i\neq \x
\end{cases}
\label{def:coherentPPM}
\eeq
The resulting association is shown in Table \ref{association}.\begin{table}
\begin{tabular}{ccccc}
Symbol & & Coherent States & & Qubits \\
\\
$\x=1$ \quad & $\Leftrightarrow$  & \quad $\ket{\alpha}\ket{0}\ket{0}\ldots\ket{0}$ \quad & $\Leftrightarrow$ & \quad $\ket{\g_1}\ket{\g_0}\ket{\g_0}\ldots\ket{\g_0}$ \\
$\x=2$ \quad & $\Leftrightarrow$  & \quad $\ket{0}\ket{\alpha}\ket{0}\ldots\ket{0}$ \quad & $\Leftrightarrow$ & \quad $\ket{\g_0}\ket{\g_1}\ket{\g_0}\ldots\ket{\g_0}$ \\
\vdots &  & \vdots & & \vdots \\
$\x=\M$ \quad &  $\Leftrightarrow$ &\quad  $\ket{0}\ket{0}\ket{0}\ldots\ket{\alpha}$ \quad & $\Leftrightarrow$ & \quad $\ket{\g_0}\ket{\g_0}\ket{\g_0}\ldots\ket{\g_1}$ \\
\end{tabular}
\caption{\label{association}Association between symbols, transmitted quantum states and qubits representation. }
\end{table}

In place of defining the transmitted quantum states in the Fock space and work with coherent states, an equivalent way to define the PPM format is to describe each slot in a qubit framework with an Hilbert space $\mathcal{H} \sim{} \mathbb{C}^2$, where either of two quantum states are possible, $\ket{\g_0}$ and $\ket{\g_1}$, corresponding to $\ket{0}$ and $\ket{\alpha}$  respectively. The representation provides the same inner product,
\beq
\inner:=\inn{\barg_i}{\barg_j}=|\inn{0}{\alpha}|^2 = \e^{-|\alpha|^2} = |\inn{\g_0}{\g_1}|^2, \quad i \neq j,
\label{def:inner}
\eeq
because from the point of view of the slot measurement, the outcome probabilities are given by the operators in the subspace spanned by $\ket{0}$ and $\ket{\alpha}$, that is isomorphic to $\mathcal{H}$.

The transmitted quantum states are then described in the tensor Hilbert space $\mathcal{H}_0 = \mathcal{H}^{\otimes \M}$, by
\beq
\ket{\g_{1,\x}}\ket{\g_{2,\x}} \ldots \ket{\g_{\M,\x}}, 
\quad \g_{i,\x} = 
\begin{cases}
\g_1 & i=\x \\
\g_0 & i\neq \x
\end{cases}
\label{def:qubitPPM}
\eeq
As we shall see later, this abstract definition allows us to focus on the \emph{consequences} of the measurements in a slot in terms of transition probabilities, rather than the actual physical implementation of the measurement. This is obtained thanks to  the possibility of implementing arbitrary projective measurement in the subspace spanned by $\{\ket{0}, \ket{\alpha}\}$, as explained in Section \ref{implementation}.

The quantum limit performance for the Pulse Position Modulation is a well known result by quantum discrimination theory \cite{Yuen1975}, and the optimal measurement operator has been characterized by means of square root measurement, exploiting the geometric uniform symmetry of the constellation \cite{Cariolaro2010}. The resulting error probability is
\begin{align}
\Pe^{\ theo} &= \frac{\M-1}{\M^2}\left( \sqrt{1+(\M-1)\inner}-\sqrt{1-\inner}\right)^2.
\label{Petheo}
\end{align}

\subsection{Quantum States and Operators in $\mathcal{H}$}

Let us define a basis $\{\ket{\X},\ket{\Y}\}$ in $\mathcal{H}$ such that without loss of generality we can write
\beq
\ket{\g_0} = \cos \theta \ket{\X} + \sin \theta \ket{\Y},\ \ket{\g_1} = \cos \theta \ket{\X} - \sin \theta \ket{\Y},
\label{statesH}
\eeq
with $\theta \in [0,\pi/4]$. The inner product \eqref{def:inner} becomes
\beq
\inner=|\inn{\g_0}{\g_1}|^2 = \cos^2 2 \theta.
\label{inner2}
\eeq

A measurement is described by a pair of POVM $(\povm_{k,\o}, \povm_{k,\l})$, each one associated with the outcome $\z_k \in \{\o, \l\}$, which must verify the completeness relation 
\beq
\povm_{k,\o} + \povm_{k,\l} = I.
\eeq
In general, the POVM can depend upon some variables, e.g. the time slot, that are indicated in the dependency upon $k$.
In addition, with the notation $\barz_k$ we indicate the sequence of outcomes $[\z_1 \dots \z_k]$.

In the binary scenario, the POVMs can be represented in the space of the Operators on $\mathcal{H}$ with the matrices
\beq
\begin{aligned}
\povm_{k,\o} &= \frac{1}{2} \left[
\begin{array}{cc}
1+a_k & b_k \\
b_k & 1-a_k
\end{array} \right], 
\\ 
\povm_{k,\l} &= \frac{1}{2} \left[
\begin{array}{cc}
1-a_k & -b_k \\
-b_k & 1+a_k
\end{array} \right],
\end{aligned}
\label{povmH}
\eeq
with $a_k^2+b_k^2 \leq 1$. In the particular case of $a_k^2+b_k^2 = 1$, the POVMs are rank-1 orthogonal projectors $(\proj_\o,\proj_1)$, and can be written without loss of generality with operators
\beq
\begin{aligned}
\povm_{k,\o} & = \proj_{k,\o} := \pure{\m_{k,\o}}, \\
\povm_{k,\l} & = \proj_{k,\l} :=  \pure{\m_{k,\l}},
\end{aligned}
\eeq
with
\beq
\begin{aligned}
\ket{\m_{k,\o}} & = \cos \ang_{k} \ket{\X} + \sin \ang_k \ket{\Y}, \\
 \ket{\m_{k,\l}} &= \sin \ang_{k} \ket{\X} - \cos \ang_k \ket{\Y},
\end{aligned}
\label{operatorH} 
\eeq
where $\ang_{k}=\frac{1}{2} \phase{a_k+\imaginary b_k}, \ \ang_k \in \left[-\frac{\pi}{2},\frac{\pi}{2} \right]$ and $\phase{\cdot}$ the four-quadrant inverse tangent \footnote{The function $\ang=\phase{a+\imaginary b}$ gives the argument of the complex number $a+\imaginary b= \sqrt{a^2+b^2} \ \e^{\imaginary \ang}$, with $\imaginary=\sqrt{-1}$ the imaginary unit.}. 


The conditional probabilities of $\z_k$ given the quantum state $\ket{\g_j}$ in the slot can be calculated from \eqref{statesH} with \eqref{povmH} as
\begin{align}
& \Prob{\z_k=i|\ \ket{\g_{k,\x}} = \ket{\g_j}}  = \p_{i|j} = \tr{\povm_{k,i} \pure{\g_j}} = \notag \\
& \quad \qquad = \frac{1+(-1)^{i+j} a_k \sin 2 \theta + (-1)^i b_k \cos 2 \theta }{2} \label{localTransPovm}
\end{align}
In the specific case of rank-1 projectors, we denote the conditional probability evaluated with \eqref{operatorH} as 
\beq
\pProj_{i|j} = |\inn{\m_{k,i}}{\g_j}|^2 = \cos^2(\theta-(-1)^j\ang_k+i\frac{\pi}{2}) \quad i,j=0,1. \label{localTrans}
\eeq
We will use the shorthand notation $\p_{i|j},\  [\p_{i,j}]$  for the conditional [joint] probability of $\z_k=i$ with the qubit $\ket{\g_{k,\x}}=\ket{\g_j}$ when it is clear from the context which random variables the realizations $i,j$ refer to. The same notation will be used with $\pProj_{i|j}\  [\pProj_{i,j}]$ in the case of rank-1 projectors employed for the measurement.

\subsection{\label{implementation}Implementation of a binary POVM pair}

In this section we state that any arbitrary POVM pair $(\povm_{k,\o},\povm_{k,\l})$ can be implemented on the subspace spanned by $\{\ket{0}, \ket{\alpha}\}$. This fact simplifies the description and the optimization of the receiver scheme for PPM signals.

It is well known \cite{Helstrom1976, Tomassoni2008, DallaPozza2013} that in the case of binary discrimination with minimum error probability, the optimal measurement operators are orthogonal projectors $(\proj'_0,\ \proj'_1)$ on the subspace spanned by the quantum states. It is then reasonable to assume
that the optimization process on the whole PPM symbol detection requires projectors (and not the more general POVMs) for the local measurements in each slot.

An optimal detection scheme for the binary coherent discrimination problem, e.g. the Dolinar receiver \cite{Dolinar1973,Cook2007}, necessarily implements such measurement projectors on the subspace spanned by $\{\ket{0},\ket{\alpha}\}$. In addition, the Dolinar receiver can implement the projectors \emph{for every possible value} of the a priori probability $\{\p'_0,\ \p'_1\}$ and coherent state parameter $\alpha$. 

Since for a given $\alpha$, each value of $\p'_0$ in the continuous range $[0,1]$ corresponds to a pair of optimal projectors $(\proj'_0,\ \proj'_1)$, there exists a version of the Dolinar receiver that from parameters $\alpha$ and $\p'_0$ implements the corresponding optimal projectors. Note that in general, $\p'_0$ does not to correspond to the a priori probability $\p_0$, but it is a \emph{fictitious} a priori probability assumed for the design of this version of the Dolinar's.


From the pairs of rank-1 projectors we can design any arbitrary projectors pairs by linear combination.
In fact, consider the spectral decomposition of the POVM pair \eqref{povmH},
\beq
\begin{aligned}
\povm_{k,\o} & = \lambda_{\newO} v_{\newO} v_{\newO}^T + \lambda_{\newL} v_{\newL} v_{\newL}^T, \\
\povm_{k,\l} & = \lambda_{\newL} v_{\newO} v_{\newO}^T + \lambda_{\newO} v_{\newL} v_{\newL}^T,
\end{aligned}
\eeq
where $\lambda_{\newO}, \lambda_{\newL}$ and $v_{\newO}, v_{\newL}$ represent the appropriate eigenvalues and eigenvectors. Any POVM pair of the type \eqref{povmH} is implemented by performing a measurement with the projectors pair 
\beq
\proj_{\newO} = v_{\newO} v_{\newO}^T, \quad \proj_{\newL} = v_{\newL} v_{\newL}^T,
\eeq
and then relabeling the outcome $\newO$ into $\o\ [\l]$ with probability $\lambda_{\newO} \ [\lambda_{\newL}]$ and the outcome $\newL$ with probability $\lambda_{\newL} \ [\lambda_{\newO}]$ respectively.

The Dolinar receiver scheme can be described both in the temporal and in the spatial domain \cite{Holevo1982,Takeoka2005}. In the former case, the scheme requires a time variant displacement operation driven by a photon counter in a feedforward fashion. In the latter case, an infinite sequence of beam splitters are required \cite{Takeoka2005}, each of them provided with a block containing a displacement operation, a photon counting and a feedforward from the previous block. 

If setup limitations or constraints, e.g. the availability of only fixed (time or spatial invariant) displacements, prevents the exact realization of a Dolinar scheme, the PPM receiver design algorithm suffers from these limitations in implementing the local measurement projectors, leading to suboptimal performances. 

In the present work we consider both the above scenarios, i.e. the availability of arbitrary local projective measurements in each slot, given for example by a Dolinar receiver, and the limitation to perform only fixed displacements. If on one hand the former solution allows greater possibilities in the design, the latter one is more practical and suitable for an experimental setup.

\section{\label{structure}Structure of an Adaptive Receiver}

The qubit description of the transmitted quantum state 
given by the tensor product of $\ket{\g_i}, \ i=0,1$ reflects the structure of the PPM signals as a temporal sequence of $M$ coherent states. Current classical receiver schemes measure the entire signal as a whole, leaving the estimation of the symbol after the measurement is concluded. However, due to the temporal sequence of the slot pulses, this measurement can be interpreted as a sequence of $M$ measurements. 

In each slot, the classical receiver performs a photon counting, also referred to as \emph{direct detection} (DD), that is an energy measurement with outcomes corresponding to the presence of any or no photons. An ideal photon counter, with unit efficiency and no dark counts, is associated to operators
\beq
\proj_\o = \pure{0},\ \proj_\l=I-\pure{0},
\label{directDetection}
\eeq
with conditional probabilities on the Fock space
\beq
\q_{i|j}=\Prob{\z_k=i| \ \ket{\coherentTensorProduct_{k,\x}} = \ket{j} }, \quad i=0,1, \ j=0,\alpha
\eeq
calculated by the Born's rule
\beq
\begin{aligned}
\q_{\o|0} & =  \tr{\proj_\o \pure{0}} = |\inn{0}{0}|^2 = 1, \\
\q_{\l|\alpha} & = \tr{(I-\pure{0}) \pure{\alpha}} = 1-\e^{-|\alpha|^2} .
\end{aligned}
\label{trans:dd}
\eeq
In the corresponding qubit framework, the measurement operators are written as
\beq
\pure{\m_{k,\o}} = \pure{\g_0}, \ \pure{\m_{k,\l}} = I - \pure{\g_0},
\label{povm:dd:qubit}
\eeq
which give the correct conditional probabilities,
\beq
\begin{aligned}
\p_{\o|0} & = |\inn{\m_\o}{\g_0}|^2 = 1, \\
\p_{\l|1} & = |\inn{\m_\l}{\g_1}|^2 = 1-\inner = 1-\e^{-|\alpha|^2} \; .
\end{aligned}
\label{trans:clas}
\eeq
The receiver estimates the transmitted symbol corresponding to the measurement where the outcome $\z_k=\l$ has been observed, that is where one or more photons have been detected. By the PPM definition, multiple outcomes equal to $\l$ are impossible to detect, while it could be possible to observe only $\o$. When all the outcomes are $\z_k=\o$, the receiver chooses at random. The error probability is thus
\beq
\Pe^{DD} = \frac{\M-1}{\M}\inner = \frac{\M-1}{\M}\e^{-|\alpha|^2}.
\label{Pe:clas}
\eeq

It is natural to ask whether it is possible to improve this receiver structure by adapting the subsequent slot measurements on the base of previous measurement outcomes. This idea was initially pointed out by Dolinar \cite{Dolinar1983} who proposed an adaptive nulling of the received signal. 

When the \emph{nulling} operation is performed in the slot, a displacement $D(-\alpha)$ is applied to the current coherent state: in the case of the ground state	 it is displaced to $\ket{-\alpha}$, while in the case of the coherent state $\ket{\alpha}$ it is displaced to $\ket{0}$. After the nulling operation, a photon counting is performed with operators \eqref{directDetection}. The result of the nulling operation is that the conditional probabilities \eqref{trans:dd} change in
\beq
\begin{aligned}
\q_{\o|0} & = |\san{0}{D(-\alpha)}{0}|^2 = \e^{-|\alpha|^2} , \\
\q_{\l|\alpha} & = \tr{(I-\pure{0}) D(-\alpha) \pure{\alpha} D(-\alpha)^\dagger} = 0.
\end{aligned}
\label{trans:nullCoherent}
\eeq
In our qubit framework, this measurement is described by the operators
\beq
\pure{\m_{k,\o}} = \pure{\g_1}, \ \pure{\m_{k,\l}} = I-\pure{\g_1}.
\eeq
with consequent conditional probabilities
\beq
\begin{aligned}
\p_{\o|0} & = |\inn{\m_{k,\o}}{\g_0}|^2 =\inner = \e^{-|\alpha|^2}, \\
\p_{\l|1} & = |\inn{\m_{k,\l}}{\g_1}|^2  = 0  \; .
\end{aligned}
\label{trans:nullQubit}
\eeq
The Dolinar receiver for PPM starts with an initial hypothesis  $\y=1$. It applies the nulling to the (unknown) coherent state in the first slot, and performs the photon counting. 

If the outcome is $\z_1=\o$, i.e. no photons are detected, the receiver believe in the temporary hypothesis $y$ and simply direct detect the remaining slots. During this stage, the receiver keeps the temporary hypothesis unless an outcome $\z_k=\l$, i.e. some photons, are observed, and in that case the hypothesis is changed in $\y=k$. 

On the contrary, if the measured outcome  in the nulled slot is $\z_1=\l$, i.e. at least one photon is recorded, the receiver completely neglects the current hypothesis, replacing it with the sequent symbol, $\y=2$. The nulling algorithm is then repeated recursively from the next slot, reducing the original problem to its $M-1$ version. 

It is easy to shown that the receiver fails in the correct discrimination of symbol $\x=i$ estimating $\y=k$ when two wrong outcomes are observed, in the nulled slot $k$ and in the slot $i$ during direct detection \cite{Dolinar1983,Guha2011}. The error probability of this scheme is
\beq
\Pe^{CN} = \frac{1}{\M} \left[ (1-\inner)^\M - 1 + \M \inner \right].
\label{Pe:nulling}
\eeq

Two architectures have been proposed \cite{Guha2011} to improve the \emph{Conditional Nulling} scheme. The key idea is that a non--exact nulling of the signal can lead to better performances, just as  the \emph{Generalized Kennedy} uses the same concept to improve the Kennedy receiver \cite{Takeoka2008}.

The first architecture, denoted in  \cite{Guha2011} as \emph{Type I}, uses the same algorithm as the PPM Dolinar receiver to choose consecutive measurements on the basis of previous outcomes, but applying a constant displacement $D(\beta)$, with $\beta \neq -\alpha$, in place of the exact nulling.
The second architecture, called \emph{Type II}, also adds a phase-sensitive amplifier with gain $G$ to squeeze the partially-nulled coherent state, further improving the performance. 
The error probability for this \emph{Improved Conditional Nulling} Type II scheme, in our ideal assumptions of unit efficiency and no dark current, becomes
\beq
\Pe^{ICN} = \frac{\q_{1|\alpha} (1 \scalebox{0.75}[1.0]{\( - \)} (1\scalebox{0.75}[1.0]{\( - \)}\q_{0|0})^{\M-1}) \scalebox{0.75}[1.0]{\( + \)} \e^{-|\alpha|^2} (\M \q_{0|0} \scalebox{0.75}[1.0]{\( - \)}1\scalebox{0.75}[1.0]{\( + \)}(1\scalebox{0.75}[1.0]{\( - \)}\q_{0|0})^\M)}{\M \q_{0|0}}
\label{Pe:Guha}
\eeq
with 
\beq
\begin{aligned}
\q_{\o|0} &= \frac{\displaystyle \textrm{exp}\left[-\frac{\displaystyle (\sqrt{G}+\sqrt{G-1})^2|\beta|^2}{\displaystyle  1+\sqrt{G-1}(\sqrt{G}+\sqrt{G-1})}\right]}{\displaystyle \sqrt{G}}, \\[10pt]
\q_{\l|\alpha} &=1-\frac{\displaystyle \textrm{exp}\left[-\frac{\displaystyle (\sqrt{G}+\sqrt{G-1})^2 |\alpha-\beta|^2}{\displaystyle  1+\sqrt{G-1}(\sqrt{G}+\sqrt{G-1})}\right]}{\displaystyle \sqrt{G}} .
\end{aligned}
\label{transitionGuha}
\eeq
Substituting $G=1$, we obtain the expression for the Type~I architecture. From the expression \eqref{Pe:Guha}, the displacement $\beta$ and the phase-sensitive amplifier $G$ have been numerically optimized to reach the maximum performance of this receiver scheme. 

In the qubit framework, the direct detection is performed again with operators \eqref{povm:dd:qubit}, while when the displacement and squeezing operations are performed, the operators are described by POVM as in \eqref{povmH}.
The POVM parameters $a_k,\ b_k$ depend upon the values of $\beta$ and $G$ employed, and can be obtained by inversion of \eqref{localTransPovm} with the transition probabilities \eqref{transitionGuha}.


The conditional nulling scheme and its improved versions share a common algorithm that creates a decision tree of the possible sequence of measurements. Proceeding from the root to the leaves, in each node it is decided which branch to take depending on the last outcome. However, it was pointed out in \cite{Guha2011} that further performance improvements can be obtained by considering different displacements $\beta_k$ for each slot $k=1 \ldots \M$. Moreover, further generalization leads to time varying displacements $\beta_k(t)$.

In addition, the decision tree of these architectures is not symmetrical, and the direct detection of all the slots after the outcome $\z_k=\l$ has been observed in a nulled slot may not be the best strategy.

\begin{figure}
	\centering
	\includegraphics{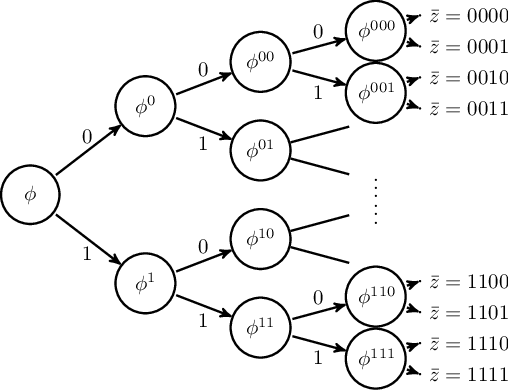}
	\caption{\label{fig:binarytree}Strategy tree for an adaptive receiver algorithm, in the case of a 4-PPM. Each node represents a measurements, identified by a parameter, and each branch a possible outcome. The parameter $\ang_k=\angDP_k(\barz_k)$ employed at the measurement $k+1$ after observing the outcomes sequence $\barz_k$ is indicated as $\ang^{\barz_k}$ to shorten the notation. The parameter defining the measurement in the first slot depends only upon the a priori information, and is denoted with $\ang$. The algorithm proceeds from the root on the left to the leaves on the right following the path indicated by the outcomes.}
\end{figure}

We propose a general structure for an adaptive receiver, where the next measurements are decided upon \emph{all} the previous outcomes $\barz_k$. The receiver algorithm defines a perfect binary tree with $\M$ levels, where each node corresponds to a slot measurement and each edge to an outcome (see Figure \ref{fig:binarytree}). 
In order to focus on the transition probabilities, we use the qubit representation to describe the measurement in the $k+1$-th slot,  i.e. after $k$ outcomes has been observed, with $\ang_k$ the parameters that specify the local measurement. For example, in the case of the projective measurements, $\ang_k$ coincides with the angle in the definition \eqref{operatorH}. We specify the function $\ang_k=\angDP_k(\barz_k)$ to define the adaptive strategy \footnote{A different notation is used in Figure \ref{fig:binarytree}, as indicated in the caption.}.

The receiver starts with the first measurement specified by $\ang_0$. Then, depending on the outcome $\z_1~=~\o$ or $\z_1~=~\l$, it proceeds with a measurement in the second slot defined by $\ang_1=\angDP_1(\o)$ or $\ang_1=\angDP_1(\l)$ respectively. In general it results $\angDP_1(\o) \neq \angDP_1(\l)$. The receiver continues to perform measurements following the path indicated by the outcomes sequence. 
After the last measurement, the final estimation is taken based on the whole outcome sequence $\barz_\M$.

This receiver structure is a generalization of the previously seen adaptive receiver. In order to achieve optimal performance, an optimization of all the parameters $\ang_k=\angDP_k(\barz_k)$, for all $k$ and $\barz_k$ is necessary.
Since the number of parameters grows exponentially in the number of levels, that is the PPM cardinality, the optimization of the final probability of correct decision is highly demanding. 
However, we can simplify the optimization problem by applying the dynamic programming algorithm, that is the topic of the next Section.


\section{\label{review}Review of Dynamic Programming}

In this Section we introduce the (discrete time) dynamic programming framework and its basic algorithm. For a more detailed review, see \cite{Bertsekas2007}.

Consider a discrete time dynamic system described by the update equation 
\beq
\s_{k+1} = f_k(\s_k, \u_k, \w_k), \quad k=0,\ldots, N-1,
\label{updateEq}
\eeq
with given initial system state $\s_0$, where 
\begin{itemize}
\item $k$ is the step index corresponding to the time.
\item $\s_k \in \S_k$ is the \emph{system state}, 
that is the collection of past information up to time $k$ useful to describe the evolution of the system and relevant for the optimization problem. To avoid misunderstanding, in the following we will use the term \emph{system state} and \emph{quantum state}, to discriminate the description of a system as in \eqref{updateEq} and the physical description given by the density operator.
\item $\u_k \in \U_k$ is the control, that is the physical variable or quantity we can use to drive the system evolution. Since we can impose the value of $\u_k$ in order to control the system, it is not described by a random variable. 
\item $\w_k \in \W_k$ is a random parameter out of our control, sometimes referred to as disturbance or noise. It can be related to $\s_k$ and $\u_k$, i.e. its probability description can depend upon $\s_k$ and $\u_k$ as in $\Prob{\cdot|\s_k, \u_k}$.
\end{itemize}

A reward function \footnote{Dynamic programming is usually formulated for a minimum optimization problem, but in our case a maximization problem is more suitable since we aim at maximize the probability of correct decision. Therefore, we introduce the concepts of reward and reward-to-go function in place of cost and cost-to-go function.} is associated with the system evolution, that in our case we can write as
\beq
g(\s_N),
\eeq
and depends upon the final system state $\s_N$. Since the evolution \eqref{updateEq} is influenced by the random variables $\w_0, \ldots, \w_{N-1},$ the final system state $\s_N$ is a random variable and the expected reward we want to maximize is 
\beq
\E[\s_N]{g(\s_N)} =  \int_{\S_N} \d \sigma \ g(\sigma)\pro_{\s_N}(\sigma),
\label{JN}
\eeq
where with the notation $\pro_r(\cdot)$ we indicate the probability density function of the random variable $r$, in this case the system state $\s_N$.

Considering the update equation \eqref{updateEq} for $k=N-1$, the expected reward can be rewritten as
\begin{align}
& \E[\s_{N-1},\w_{N-1}]{g(f_{N-1}(\s_{N-1}, \u_{N-1}, \w_{N-1}))} = \notag \\
& \qquad \qquad = \int\limits_{\S_{N-1}} \d \rho \int\limits_{\W_{N-1}} \d \omega \ g(f_{N-1}(\rho, \u_{N-1},\omega)) \notag \\ 
& \qquad \qquad \qquad \qquad \qquad \qquad \times \pro_{\s_{N-1},\w_{N-1}}(\rho, \omega).
\label{JN2}
\end{align}


By \eqref{updateEq} we can iterate the substitutions backward in the index $k$ to obtain a reformulation of the expected reward \eqref{JN} in terms of the controls and the initial  system state $\s_0$, 
\begin{align}
& \int\limits_{\W_0} \d \omega_0 \ \cdots \int\limits_{\W_{N-1}} \d \omega_{N-1} \ \pro_{\w_0, \ldots, \w_{N-1}} (\omega_0, \ldots, \omega_{N-1}) \notag \\
& \qquad \times  g(f_{N-1}(\ldots f_0(\s_0,\u_0,\omega_0), \ldots , \u_{N-1},\omega_{N-1})) 
\label{Jo}
\end{align}
where we use the fact that the initial state $\s_0$ is known, and we have explicitly indicated the composition of the update functions $f_{N-1}, f_{N-2}, \ldots, f_0$ from $\s_0$ with the variables $\u_0, \ldots, \u_{N-1},$ $\w_0=\omega_0, \ldots, \w_{N-1}=\omega_{N-1}$. 

In seeking the maximization of \eqref{Jo}, we can employ different strategies. For example, the values of the control $\u_0, \ldots, \u_{N-1}$ can be determined before the system starts, and then applied during the evolution, or we can postpone the choice of $\u_k$ at time $k$ since there's no penalties in delaying the decision. In particular, this latter strategy allows to define $\u_k$ as a \emph{function} of the system state $\s_k$, 
\beq
\u_k = \pi_k(\s_k), \quad \s_k \in \S_k,\  \u_k \in \U_k,
\label{adaptiveMap}
\eeq
leading to an adaptive control algorithm. Its performance are not worse than the fixed control, and we can take advantage of the information gained from time $0$ to $k$.

The set of functions $\bar{\pi}=\left(\pi_0, \pi_1, \ldots, \pi_{N-1} \right)$ is called a \emph{policy}. We can define the \emph{reward-to-go} function at time $k$ from the current system state $\s_k$ 
as the function
\beq
\J_k: \S_k \times \U_k \times \Phi_{k+1} \times \ldots \times \Phi_{N-1} \longmapsto \mathbb{R}
\eeq
specified as
\begin{align}
&\J_k \left(\rho, \nu ,\pi_{k+1} , \ldots , \pi_{N-1} \right) = \notag \\
& \int\limits_{\W_k} \d \w_k \ \cdots \mkern-10mu \int\limits_{\W_{N-1}} \mkern-10mu \d \w_{N-1} \ \pro_{\w_k, \ldots, \w_{N-1}|\s_k} (\omega_k, \ldots, \omega_{N-1}|\rho)  \notag \\
& \qquad \qquad  \times g(f_{N-1}(\ldots f_k(\rho,\nu,\omega_k), \ldots , \u_{N-1},\omega_{N-1}))
\label{def:Jk}
\end{align}
with $\Phi_k=\U_k^{\S_k}$ the set of all possible functions $\pi_k~:~\S_k~\mapsto~\U_k$.

Define $\bar{\pi}^\ast=\left(\pi_0^\ast, \pi_1^\ast, \ldots, \pi_{N-1}^\ast\right)$ the optimal policy, that is the one that maximize $\J_0$
\beq
\bar{\pi}^\ast := \operatorname*{argmax}_{\bar{\pi}} \ \J_0(\s_0,\bar{\pi}(\s_0)),
\eeq
and define 
\beq
\J_0^\ast(\s_0) := \J_0(\s_0, \bar{\pi}^\ast(\s_0))
\eeq
the optimal reward from $\s_0$.

The dynamic programming algorithm relies on the following idea. 

\vspace{2mm}
\noindent {\bf Principle of Optimality \cite{Bertsekas2007}} \\
\emph{Let $\bar{\pi}^\ast=\left( \pi_0^\ast, \pi_1^\ast, \ldots, \pi_{N-1}^\ast \right)$ be the optimal policy that maximizes the reward $\J_0$ and let $\s_1, \s_2, \ldots, \s_N$ be the corresponding system state evolution. Consider the subproblem of the maximization of the reward-to-go function from time $k$ with $\s_k=\sigma$, 
\beq
\operatorname*{max}_{\pi_k, \ldots, \pi_{N-1}} \ \J_k(\sigma,\pi_k(\sigma),\ldots ,\pi_{N-1}) =: \J_k^\ast(\sigma).
\label{subproblem}
\eeq
The optimal policy for this subproblem is the truncated sequence $\left( \pi_k^\ast, \pi_{k+1}^\ast, \ldots, \pi_{N-1}^\ast \right)$.
}
\vspace{2mm}

The maximization of $\J_0$ with respect to the policy $\left( \pi_0, \ldots, \pi_{N-1} \right)$ with multivariate calculus requires the solution of a complicated system with equations in all the variables $\pi_k$. Instead, the dynamic programming algorithm decomposes the main problem into a sequence of subproblems.

\vspace{2mm}
\noindent {\bf Dynamic Programming Algorithm \cite{Bertsekas2007}} \\
\emph{The optimal reward $\J_0^\ast$ is the last step of the following algorithm, which proceeds backwards from $N$ to $0$, 
\begin{itemize}
\item[1.] define the initial condition 
\beq
\J_N^\ast(\sigma) = g(\sigma) , \quad \sigma \in \S_N
\eeq
\item[2.] for $k=N-1, \ldots, 0$, for all $\rho \in \S_k$  evaluate the optimal control and the optimal reward-to-go function at time $k$, namely
\begin{align}
\pi_k^\ast(\rho)  & = \operatorname*{argmax}_{\nu \in \U_k} \ \J_k(\rho,\nu,\pi_{k+1}^\ast,  \ldots , \pi_{N-1}^\ast) \\
& = \operatorname*{argmax}_{\nu \in \U_k} \ \E[\w_k] {\J_{k+1}^\ast(f_k(\rho,\nu,\w_k))} , \\
\J_k^\ast(\rho) & = \J_k\left( \rho,\pi_k^\ast(\rho),\pi_{k+1}^\ast, \ldots, \pi_{N-1}^\ast\right ).
\end{align}
\item[3.] the optimal reward and the optimal policy are
\begin{align}
\J_0^\ast(\s_0) & = \J_0(\s_0,\pi_0^\ast(\s_0), \ldots, \pi_{N-1}^\ast ),\\[6pt]
\bar{\pi}^\ast&=\left( \pi_0^\ast, \pi_1^\ast, \ldots, \pi_{N-1}^\ast \right).
\end{align}
\end{itemize}
}
\vspace{2mm}

At each step $k$, assuming to know by induction the optimal policy  $\left(\pi_{k+1}^\ast, \ldots, \pi_{N-1}^\ast\right)$ and the optimal reward $\J_{k+1}^\ast(\cdot)$,
the algorithm considers $\s_k=\rho$ as the initial system state for the time evolution from $k$ to $N$, and maximize with respect to the control $\u_k=\nu$. This maximization is solved for every possible $\s_k \in \S_k$, in order to define the function $\pi_k^\ast(\s_k)$ and the optimal reward $\J_{k}^\ast(\s_{k})$. This step is repeated until $k=0$. Note that at each step, only one control variable $\u_k$ is involved in the maximization, simplifying its optimization.

\subsection{\label{reformulation}Reformulation of the Dynamic Programming Algorithm}
In some optimization problems, it could be of interest to include the probability of the actual system state $\s_k$ among the information to pass from an iteration to another, as a component of the system state. In this case, the probability of the system state $\pro_{\s_k}(\cdot)$ refers to the other components. As we will see later, this allows us to reduce the size of the system state thus speeding up the computation of the algorithm.

A consequence of this new definition is that in the steps 1. and 2. of the dynamic programming algorithm must be evaluated assuming a possible value for the probability of the state. This is not a problem since even in the original formulation the system state probability $\pro_{\s_k}(\cdot)$ is not known, but it is obtained by the backward steps of the algorithm. In the same manner, non-admissible values of the system state probabilities are excluded by the backward procedure.

Including the system state probability allows us to define a reward-to-go function which uses its value, with joint probabilities rather than conditional probabilities, as in 
\begin{align}
&\Jc_k \left(\rho, \nu ,\pi_{k+1} , \ldots , \pi_{N-1} \right) = \notag \\
& \int\limits_{\W_k} \d \w_k \ \cdots \mkern-10mu \int\limits_{\W_{N-1}} \mkern-10mu \d \w_{N-1} \ \pro_{\s_k,\w_k, \ldots, \w_{N-1}} (\rho,\omega_k, \ldots, \omega_{N-1})  \notag \\
& \qquad \qquad  \times g(f_{N-1}(\ldots f_k(\rho,\u_k,\omega_k), \ldots , \u_{N-1},\omega_{N-1}))\\
& = \pro_{\s_k} (\rho) \int\limits_{\W_k} \d \w_k \ \cdots \int\limits_{\W_{N-1}}\d \w_{N-1} \notag\\
& \qquad \qquad \times  g(f_{N-1}(\ldots f_k(\rho,\u_k,\omega_k), \ldots , \u_{N-1},\omega_{N-1}))  \notag \\
& \qquad \qquad \qquad \times \pro_{\w_k, \ldots, \w_{N-1}|\s_k} (\omega_k, \ldots, \omega_{N-1}|\rho),
\label{def:Jk2} \\
& = \pro_{\s_k} (\rho) \J_k \left(\rho, \nu ,\pi_{k+1} , \ldots , \pi_{N-1} \right)
\end{align}

In addition, the dynamic programming algorithm can be reformulated with integrals rather that expectations, as in the following procedure. 

\vspace{2mm}
\noindent {\bf Dynamic Programming Algorithm (Reformulation)} \\
\emph{The optimal reward $\J_0^\ast$ is the last step of the following algorithm, which proceeds backwards from $N$ to $0$, 
\begin{itemize}
\item[1.] define the initial condition 
\beq
\Jc_N^\ast(\sigma) = \tilde{g}(\sigma) := g(\sigma) \pro_{\s_N}(\sigma) , \quad \sigma \in \S_N
\eeq
\item[2.] for $k=N-1, \ldots, 0$, for all $\rho \in \S_k$  evaluate the optimal control and the optimal reward-to-go function at time $k$, namely
\begin{align}
\pi_k^\ast(\rho)  & = \operatorname*{argmax}_{\nu \in \U_k} \ \Jc_k(\rho,\nu,\pi_{k+1}^\ast,  \ldots , \pi_{N-1}^\ast) \\
& = \operatorname*{argmax}_{\nu \in \U_k} \ \int\limits_{\W_{k}} \d \omega_k \ \Jc_{k+1}^\ast(f_k(\rho,\nu,\omega_k)) , \\
\Jc_k^\ast(\rho) & = \Jc_k\left( \rho,\pi_k^\ast(\rho),\pi_{k+1}^\ast, \ldots, \pi_{N-1}^\ast\right ).
\end{align}
\item[3.] the optimal reward and the optimal policy are
\begin{align}
\J_0^\ast(\s_0) & = \Jc_0(\s_0,\pi_0^\ast(\s_0), \ldots, \pi_{N-1}^\ast ),\\[6pt]
\bar{\pi}^\ast&=\left( \pi_0^\ast, \pi_1^\ast, \ldots, \pi_{N-1}^\ast \right).
\end{align}
\end{itemize}
}
\vspace{2mm}
Note that since $\J_k$ and $\Jc_k$ differs by only the factor $\pro_{\s_k}$, the values of $\pi_k^\ast(\rho)$ maximizing the two definitions of the reward-to-go function are the same.

\section{\label{optimization}Optimization algorithm}

In this Section we follow the dynamic programming algorithm to optimize the parameters tree depicted in Figure \ref{fig:binarytree}. More details are reported in Appendix \ref{preliminar}, where some preliminary lemmas are explained in order to better understand the algorithm.

\subsection{System state of the Algorithm}

The dynamic programming algorithm applies to dynamic systems, whose time evolution is described by a system of equations involving its system state. In our case, the time evolution occurs in discrete time steps.

We refer to the iteration 
$k$ of the algorithm as the moment just \emph{after} the sequence $\barz_k$ has been observed, that is \emph{after} the $k$-th measurement and \emph{before} the $k+1$-th one. When $k=\M$, no more measurements are performed but a final estimation is made in order to choose $\y$. By extension, we can define $k=0$ as the time step before the measurement in the first slot.

In order to define the system state $\s_k$, we need to evaluate the information that is necessary to express the probability of correct decision. In general, a receiver strategy considers first a measurement stage, where information about the system under investigation are acquired performing measurement on it, and later an estimation stage, where  the information are processed in order to formulate the answer to our detection problem.

The estimation stage is a (possibly probabilistic) map $\f$ that assigns to each measurement outcome sequences $\barz_\M$ an estimate $\y$ of the transmitted symbol $\x$. 
The final probability of correct decision is therefore rewritten as
\begin{align}
\Pc & = \Prob{\y = \x} = \sum_{i=1}^{\M} \Prob{\y=i, \x=i}\notag \\
& = \sum_{i=1}^{\M} \sum_{\zM \in \mathcal{Z}_{\M}} \Prob{\y=i, \x=i, \barz_{\M} = \zM} \notag \\
& = \sum_{i=1}^{\M} \sum_{\zM \in \mathcal{Z}_{\M}} \Prob{\y=i| \x=i, \barz_{\M} = \zM} \p_{i,\zM}.  \label{estimationStage}
\end{align}
where $\Prob{\x=i, \barz_{\M} = \zM}=\p_{i,\zM}$.
%

The performance index \eqref{estimationStage} is maximized by a deterministic map $\f$ according to the maximum a posteriori (MAP) criterion, 
\beq
\f(\zM) = \operatorname*{argmax}_{i \in \{1, \dots, \M\}} \ \p_{i,\zM},
\label{MAPhx}
\eeq
such that the probability of correct decision reads
\begin{align}
\Pc & = \sum_{\zM \in \mathcal{Z}_\M}  \Prob{\x=\f(\zM), \barz_{\M} = \zM} = \sum_{\zM \in \mathcal{Z}_\M} \operatorname*{max}_{i\leq\M} \ \p_{i,\zM} \label{Pclast} 
\end{align}
The probability of correct decision \eqref{Pclast} corresponds to the expected reward function \eqref{Jo}, namely
\beq
\E[\w_0, \ldots, \w_{N-1}]{g(\s_\M)} = \sum_{\zM \in \mathcal{Z}_\M} \left( \operatorname*{max}_{i\leq\M} \ \p_{i|\zM}  \right) \p_{\zM},
\eeq
and therefore we can identify $\w_k = \z_{k+1}$ and 
\beq
g(\s_\M(\zM)) = \operatorname*{max}_{i\leq\M} \ \p_{i|\zM}.
\eeq 

While the estimation stage is completely optimized by \eqref{MAPhx}, in order to optimize the measurement stage, it seems natural to consider 
the information described by the vector of the conditional probabilities given the outcome $\bar{z}_k$,
\beq
\label{conditionalProbabilityVector}
\barp_{\barz_k} = \left[
\begin{array}{c}
\p_{1|\bar{z}_k} \\
\p_{2|\bar{z}_k} \\
 \vdots \\
 \p_{M|\bar{z}_k}
\end{array}
\right].
\eeq

However, not all the entries in \eqref{conditionalProbabilityVector} are necessary in order to solve \eqref{MAPhx} and therefore define the \emph{system state} of the receiver algorithm. In fact, given the outcome sequence $\barz_k$, due to the local binary discrimination in each slot it turns out that only two conditional probabilities are necessary, $\p_{\M|\barz_k}$ and $\p_{\mx|\barz_k}$, with $\mx(\barz_k) $ the maximum a posteriori (MAP) estimate of $\x$ among the symbols $1, \ldots, k$, i.e.
\beq
\mx(\barz_k) = \operatorname*{argmax}_{i \leq k} \ \p_{i|\barz_k}. \label{maxStepk}
\eeq

These considerations lead to a definition of the state as in 
\beq
\s_k(\barz_k)=(\p_{\mx|\barz_k},\p_{\M|\barz_k}, \mx(\barz_k), \p_{\barz_k}).
\label{def:state2}
\eeq 
where we include in the last component the probability of the state, as explained in the previous Section \ref{reformulation}.
However, due to the reformulation of the dynamic programming algorithm, it suffices to define as a system state $\s_k \in \S_k$ the triple
\begin{align}
\s_k(\barz_k) & =   (\p_{\mx,\barz_k},\p_{\M,\barz_k}, \mx(\barz_k)),  \label{def:state}  \\[4pt]
\S_k \ \ & =  \{(u,v, i): 0\leq u \leq 1,0\leq v \leq 1,  \notag \\
&  \qquad \qquad u+v \leq 1, i \in \{1, \ldots, k \}\}, \label{def:stateSpace}
\end{align}
with joint probabilities rather than conditional ones, and use the definition of probability of correct decision \eqref{Pclast}. This definition is not equivalent to \eqref{def:state2}, since we cannot recover \eqref{def:state2} from \eqref{def:state}, but still it suffices to perform the optimization. Note that $\mx(\barz_k)$ can be defined in the same manner of \eqref{maxStepk} with joint probabilities.

We highlight that the system state $\s_k$ is a random variable, that depends upon the realization of the outcomes sequence $\barz_k$. However, to shorten the notation, when assuming a given $\barz_k$, we drop its dependency.

We show that by definition \eqref{def:state} we can describe the evolution of the system state with an update equation. Later, in the next Section, we show that we can write the probability of correct decision as a function of $\s_\M(\barz_\M)$.

Consider the system state $\s_k(\barz_k)$ and the outcome sequences that can be generated from $\barz_k$ with $\z_{k+1}=\o$ or $\z_{k+1}=\l$, employing the parameters $\ang_k = \angDP_k(\barz_k)$ in the measurement.
The joint probabilities $\p_{\M, [\barz_k \o]}, \p_{\M, [\barz_k \l]}$ can be easily obtained from the transition probabilities in equations \eqref{localTrans}, in the case $k<\M-1$ by
\beq
\begin{array}{rcl}
\p_{\M, [\barz_k \o]} & = & \pProj_{\o|0} \p_{\M, \barz_k},  \\
\p_{\M, [\barz_k \l]} & = & \pProj_{\l|0} \p_{\M, \barz_k}, 
\end{array}
\label{jointK}
\eeq
while if $k=\M-1$ by
\beq
\begin{array}{rcl}
\p_{\M, [\barz_k \o]} & = & \pProj_{\o|1} \p_{\M, \barz_k},  \\
\p_{\M, [\barz_k \l]} & = & \pProj_{\l|1} \p_{\M, \barz_k}.
\end{array}
\label{jointM}
\eeq
Note that given the system state $\s_k(\barz_k)$, the probability $\p_{\M,{\barz_k}}$ equals the joint probability 
\beq
\p_{i,\barz_k}=\Prob{\x=i,\barz_k}, \quad i=k+1, k+2, \ldots, \M
\nonumber
\eeq
due to the equal a priori probability and the same quantum states $\ket{\g_0}$ in the first $k$ slots for all symbols $\x=k+1, \ldots, \M$ (see Lemma \ref{probabilityM} in Appendix \ref{preliminar}).

In the case of $\p_{\mx(\barz_k \o),[\barz_k \o]}$ and $\p_{\mx(\barz_k \l),[\barz_k \l]}$, in order to obtain the update equation we apply its definition,
\begin{align}
& \p_{\mx(\barz_k \o),[\barz_k \o]} = \operatorname*{\max}_{i \leq k+1} \ \{\p_{i,[\barz_k \o]} \} \notag \\
& \qquad = \max \ \{ \pProj_{\o|1} \p_{k+1,\barz_k}, \ \operatorname*{max}_{i\leq k} \ \{ \pProj_{\o|0} \p_{i,\barz_k} \} \} \notag \\
& \qquad = \max \ \{ \pProj_{\o|1} \p_{\M,\barz_k}, \ \pProj_{\o|0} \p_{\mx(\barz_k),\barz_k} \} \label{argmax0},\\
& \p_{\mx(\barz_k \l),[\barz_k \l]} = \operatorname*{\max}_{i \leq k+1} \ \{\p_{i,[\barz_k \l]} \} \notag \\
& \qquad = \max \ \{ \pProj_{\l|1} \p_{k+1,\barz_k}, \ \operatorname*{max}_{i\leq k} \ \{ \pProj_{\l|0}\p_{i,\barz_k} \} \} \notag \\
& \qquad = \max \ \{ \pProj_{\l|1} \p_{\M,\barz_k}, \ \pProj_{\l|0} \p_{\mx(\barz_k),\barz_k} \} \label{argmax1}.
\end{align}
Thereby, the symbol $\mx(\barz_k)$ is updated into
\beq
\mx([\barz_k \ \z_{k+1}]) \in \{\mx(\barz_k),k+1\}
\label{updateMx}
\eeq
accordingly with the term maximizing \eqref{argmax0} and \eqref{argmax1}. This means that at each update of the system state, the symbol $\mx(\barz_k)$ can be replaced only by the symbol corresponding to the current slot. 

Given the update equations, we have to specify the initial system state $\s_0$ of the algorithm before the first measurement, that does not depend upon any outcomes, $\barz_0=\varnothing$. This system state collects the a priori information, and due to the equal a priori distribution for the symbols, we can define
\beq
\s_0 = \left ( \frac{1}{\M},\frac{1}{\M}, \cdot \right ),
\eeq
where it is unnecessary to specify $\mx(\varnothing)$, since the first update of the system state leads to the trivial maximization
\beq
\mx(\z_1)  =  \operatorname*{argmax}_{i \leq 1} \ \p_{i,\z_1} = 1.
\eeq

The parameters $\ang_k=\angDP_k(\barz_k)$ appear (implicitly) in the transition probabilities of update equations \eqref{jointK}-\eqref{updateMx}, highlighting their role of control variables.
In the next Section we formulate the corresponding expected reward function, and following the dynamic programming we define $\ang_k$ as function of the state $\s_k$,
\beq
\ang_k = \angDP_k(\s_k(\barz_k))
\eeq
optimizing for each $\barz_k$.

\subsection{Reward-to-go function}

In this section we rewrite the probability of correct decision as a function of the system state  and find a suitable definition in terms of the reward-to-go functions.

Consider now the system state $\s_k(\zM)$ before the $k+1$-th measurement, and define the reward-to-go function
\begin{align}
& \fTable_{k}(\s_k(\zM), \ang_k, \angDP_{k+1}, \ldots, \angDP_{\M-1}) = \notag \\
& \qquad = \sum_{\zM' \in \mathcal{Z}_{\M-k}} \p_{\mx([\zM \ \zM']),[\zM \ \zM']}
\label{def:fTable}
\end{align}
where the set $\mathcal{Z}_{\M-k}$ contains all the possible sequences $[\z_{k+1} \ldots \z_\M]$ composed by $\M-k$ outcomes. The dependency upon the variables $\s_k(\zM),\ang_k, \angDP_{k+1}, \ldots, \angDP_{\M-1}$ in the RHS of \eqref{def:fTable} is implicit in the transition probabilities of the outcomes.

It is trivial to see that
\begin{align}
&\fTable_{0}(\s_0, \angDP_0(\s_0), \ldots, \angDP_{M-1})  = \Pc
\end{align}
and
\begin{align}
& \fTable_{\M-1}(\s_{\M-1}(\zM), \ang_{\M-1}) = \sum_{\z_\M=\o,\l} \p_{\mx([\zM \ \z_{\M}]),[\zM \ \z_{\M}]} (\ang_{\M-1})\notag \\[6pt]
& \qquad = \p_{\mx([\zM \ \o]),[\zM \ \o]}(\ang_{\M-1}) + \p_{\mx([\zM \ \l]),[\zM \ \l]} (\ang_{\M-1})\notag \\[6pt]
& \qquad  = \Prob{\z_\M = \o| \x=i} \p_{i, \zM } + \Prob{\z_\M = \l| \x=j}\p_{j, \zM},
\label{lastStep}
\end{align}
with $i=\f([\zM \ \o])$ and $j=\f([\zM \ \l])$.

We consider for the moment the possibility to perform arbitrary local projective measurements, as explained in Section \ref{implementation}. Therefore, the transition probabilities
$\Prob{\z_\M =\z| \x=i} = \pProj_{\z|i}$ can be evaluated as in \eqref{localTrans}.

As explained in the Lemma \ref{discrimination2} in Appendix \ref{preliminar} , expression \eqref{lastStep} is the probability of correct decision of the binary discrimination problem between symbols $i=\mx(\zM)$ and $j=\M$, with the joint probabilities $\p_{i, \zM },\p_{j, \zM }$ provided by the system state $\s_{\M-1}(\zM)$. The expression admits the analytical maximixation
\beq
\fTable_{\M-1}^\ast(\s_{\M-1}) = \frac{1}{2} \left [ \p_{\mx} +  \p_{\M} + \sqrt{(\p_{\mx} +  \p_{\M})^2-4 \p_{\mx} \p_{\M} \inner} \right ] \label{initialfTable}
\eeq
obtained employing the angle in the $\M$-th measurement
\beq
\angDP_{\M-1}^\ast(\s_{\M-1}) = \frac{1}{2} \phase{\cos 2 \theta (\p_\mx - \p_\M) +\imaginary \sin 2 \theta (\p_\mx + \p_\M)} \label{initialPhi}
\eeq
where in both \eqref{initialfTable} and \eqref{initialPhi} we drop the dependency from $\barz_{\M-1}$.

Moreover, we can easily write down the update equation for the reward-to-go
\begin{align}
& \fTable_k(\s_k(\zM), \ang_k,\angDP_{k+1},\ldots, \angDP_{\M-1})  = \sum_{\substack{\zM' \in \\ \mathcal{Z}_{\M-k}}} \p_{\mx([\zM \ \zM']),[\zM \ \zM']} \notag \\
& \qquad = \sum_{\z_{k+1}} \sum_{\substack{\zM'' \in \\ \mathcal{Z}_{\M-k-1}}} \p_{\mx([\zM \ \z_{k+1} \ \zM'']),[\zM \ \z_{k+1} \ \zM'']} \notag \\
& \qquad  =  \fTable_{k+1}(\s_{k+1}([\zM \ \o]),\angDP_{k+1},\angDP_{k+2},\ldots, \angDP_{\M-1}) \notag \\[4pt]
& \qquad \qquad +  \fTable_{k+1}(\s_{k+1}([\zM \ \l]),\angDP_{k+1},\angDP_{k+2},\ldots, \angDP_{\M-1}) \label{update1}
\end{align}
In equation \eqref{update1}, the role of $\ang_k$ comes into the update of the system state. In fact, for each outcome $\z_{k+1}$, two evolutions of $\s_k([\zM \ \z_{k+1}])$ are possible, which depend on parameters $\ang=\ang_k$ that appear in the transition probabilities of \eqref{argmax0} and \eqref{argmax1}. Therefore, four possible evolutions must be considered in evaluating \eqref{update1}, indicated with A, B, C and D in equations \eqref{AC}-\eqref{BD}. Since we want to maximize the probability of correct decision, $\fTable_k^\ast(\s_k)$ is the maximum between these possibilities, as in \eqref{updatefTable}.

\begin{figure*}
\begin{widetext}
\begin{align}
\fTable_{k,A}(\s_k,\ang) &= \fTable_{k+1}^\ast( \ {\overbrace{ \pProj_{\o|0}(\ang) \p_\mx,\ \pProj_{\o|0}(\ang) \p_\M,\mx}^{\displaystyle \s_{k+1}([\barz_k \o])}} \ )  +  \fTable_{k+1}^\ast(\ {\overbrace{\pProj_{\l|0}(\ang) \p_\mx ,\ \pProj_{\l|0}(\ang) \p_\M,\mx}^{\displaystyle \s_{k+1}([\barz_k \l])}}\ ),\label{AC}\\
\fTable_{k,B}(\s_k,\ang) &=\fTable_{k+1}^\ast(\ \pProj_{\o|0}(\ang) \p_\mx,\ \pProj_{\o|0}(\ang) \p_\M,\mx \ ) + \fTable_{k+1}^\ast(\ \pProj_{\l|1}(\ang) \p_\M ,\ \pProj_{\l|0}(\ang) \p_\M,k+1 \ ),  \\ 
\fTable_{k,C}(\s_k,\ang) &=\fTable_{k+1}^\ast(\ \pProj_{\o|1}(\ang) \p_\M,\ \pProj_{\o|0}(\ang) \p_\M,k+1 \ ) + \fTable_{k+1}^\ast(\ \pProj_{\l|0}(\ang) \p_\mx ,\ \pProj_{\l|0}(\ang) \p_\M, \mx \ ),  \\
\fTable_{k,D}(\s_k,\ang) &= \fTable_{k+1}^\ast(\ \pProj_{\o|1}(\ang) \p_\M,\ \pProj_{\o|0}(\ang) \p_\M,k+1 \ ) + \fTable_{k+1}^\ast(\ \pProj_{\l|1}(\ang) \p_\M ,\ \pProj_{\l|0}(\ang) \p_\M,k+1 \ ) \label{BD}
\\[6pt]
\fTable_k^\ast(\s_k) \ \ &  = \operatorname*{\max}_{\ang} \ \{ \fTable_{k,A}(\s_k,\ang), \fTable_{k,B}(\s_k,\ang), \fTable_{k,C}(\s_k,\ang), \fTable_{k,D}(\s_k,\ang)\} \label{updatefTable}\\
\angDP_k^\ast(\s_k) \ \ &=	 \operatorname*{argmax}_{\ang} \ \{ \fTable_{k,AC}(\ang), \fTable_{k,AD}(\ang), \fTable_{k,BC}(\ang), \fTable_{k,BD}(\ang)\} \label{angK}
\end{align}
\end{widetext}
\end{figure*}

Along with the reward-to-go function, we define the function $\angDP_k^\ast$ that represents the optimal value of the control variable corresponding to the current system state, $\ang_k = \angDP_k^\ast(\s_k)$, to employ in the measurement $k+1$ given that the outcome sequence $\barz_k$ has been observed.

\subsection{Dynamic Programming Algorithm}

In the dynamic programming algorithm, we have to evaluate the reward-to-go function at iteration $k$ for each possible values of the system state $\s_k$. The expression of $\fTable_k$ depends on the variables $\p_\mx,\ \p_\M$ and not upon the particular sequence $\barz_k$. Therefore, in the following we drop the dependence of the system state from $\barz_k$.

The optimization algorithm used to evaluate the performance of the adaptive receiver can be summarized by the following step by step procedure:
\begin{itemize}
\item[1.] Evaluate the reward-to-go function $\fTable_{\M-1}^\ast$ and the 
function $\angDP_{\M-1}^\ast$ for each $(\p_\mx,\p_\M), \ \p_\mx~+~\p_\M~\leq~1$ as in \eqref{initialfTable} and \eqref{initialPhi}.
\item[2.] Repeat step 3. and 4. for $k=\M-1, \ldots, 1$.
\item[3.] From $\fTable_{k+1}^\ast$, for each $(\p_\mx,\p_\M), \ \p_\mx+\p_\M \leq 1$  evaluate $\fTable_k^\ast$ and $\angDP_k^\ast$ as in \eqref{updatefTable} and \eqref{angK} respectively.
\item[4.] For each $(\p_\mx,\p_\M), \ \p_\mx+\p_\M \leq 1$, depending on the association A, B, C or D of \eqref{AC}-\eqref{BD} used in the previous step, define the children nodes of $\s_{k}(\z_k)$ generated with outcome $\z_{k+1}=\o$ and $\z_{k+1}=\l$
\beq
\nextS(\s_{k})=\{\s_{k+1}([\barz_{k} \o]),\ \s_{k+1}([\barz_{k}\l])\}
\eeq
Note that in $\s_{k+1}([\barz_{k} \o])$ and in $\s_{k+1}([\barz_{k}\l])$ we can define $\mx([\barz_{k}\o])$ and $\mx([\barz_{k}\l])$ only in the case it is equal to $k+1$, while in the case $\mx([\barz_{k} \z_{k+1}]) = \mx(\barz_{k})$ we cannot assign an exact value, because $\mx(\barz_{k}) \in \{1, \ldots, k\}$. Instead, we can assign the label ``previous'' indicating the value is $\mx \leq k$, that will be defined in later iterations of the optimization.
\item[5.] Evaluate the parameter in the first measurement and the performances of the adaptive algorithm from $\s_0$ as 
\begin{align}
\angDP_0^\ast & = \operatorname*{argmax}_\ang \ \fTable_1^\ast\left(\frac{\pProj_{\o|1}(\ang)}{\M}, \frac{\pProj_{\o|0}(\ang)}{\M}, 1 \right) \notag \\
& \qquad \qquad \qquad + \fTable_1^\ast\left(\frac{\pProj_{\l|1}(\ang)}{\M}, \frac{\pProj_{\l|0}(\ang)}{\M}, 1 \right) \label{ang1} \\
\Pc &= \fTable_1^\ast\left(\frac{\pProj_{\o|1}(\angDP_0^\ast)}{\M}, \frac{\pProj_{\o|0}(\angDP_0^\ast)}{\M}, 1 \right) \notag \\
& \qquad \qquad  + \fTable_1^\ast\left(\frac{\pProj_{\l|1}(\angDP_0^\ast)}{\M}, \frac{\pProj_{\l|0}(\angDP_0^\ast)}{\M}, 1 \right) \label{fTable0}
\end{align}
\end{itemize}

In order to reconstruct the binary tree parameters and find the estimation region, we need to retrace the optimization steps forward. In the following procedure, two binary trees are built, one with nodes the system states $\s_{k}$ and the other with nodes corresponding to the parameter $\ang_k$. 
The levels $k=0, 1, \ldots, \M-1$ of the trees represent the system state before $k+1$-th measurement, the  edges between the nodes correspond to a measurement outcome $\z_k=\o$ or $\z_k=\l$. The path from the root to the node gives the outcomes sequence.
	
In particular, retracing the path of the binary tree we can fill up the system state substituting the labels ``previous'' with the correct symbol $\mx(\barz_k)$.

The construction of the binary trees is given by the following steps:
\begin{itemize}
\item[6.] Define the initial system state $\s_\varnothing$ as the root of the binary tree of the system states.
\item[7.] Define $\ang_0=\angDP_0^\ast(\s_0)$ as the root of the tree of the parameters.
\item[8.] Define the children nodes of the system state $\s_0$, the one corresponding to the outcome $\z_1=\o$,
\beq
\s_1(\o) = \left(\frac{\pProj_{\o|1}(\angDP_0^\ast)}{\M},\frac{\pProj_{\o|0}(\angDP_0^\ast)}{\M}, 1 \right),
\eeq
and the other corresponding to $\z_1=\l$,
\beq
\s_1(\l) = \left(\frac{\pProj_{\l|1}(\angDP_0^\ast)}{\M},\frac{\pProj_{\l|0}(\angDP_0^\ast)}{\M}, 1 \right).
\eeq
\item[9.] Repeat step 10. for $k=2, \ldots, \M-1$.
\item[10.] For each node $\s_k(\barz_k)$ of the level $k$ in the binary parameters tree, the parameter corresponding to the \emph{next} measurement is
\beq
\angDP_k^\ast(\s_k(\barz_k))
\eeq
and in the next level of the system state tree add  $\s_{k+1}([\barz_k \o])$ and $\s_{k+1}([\barz_k \l])$, replacing, if present, the label ``previous'' with the symbol $\mx(\s_k(\barz_k))$.
\end{itemize}
Once completed these trees, following the outcome sequence through the parameter tree we get the parameter $\ang_k=\angDP_k(\s_k(\barz_k))
$ to employ in the $k+1$-th measurement. The region of estimation are defined in this way: if the sequence ends in $\z_\M=\l$, it is attributed to $\y=\M$, otherwise for $\z_\M=\o$ it is assigned to $\y=\mx(\s_{\M-1}(\barz_{\M-1}))$.

\subsection{\label{adaptive} Adaptive Receiver Algorithm with Suboptimal Local Measurements}

The adaptive receiver algorithm we proposed in Section \ref{structure} bases its performance on two key ingredients, the possibility of implementing arbitrary local projective measurements and the dynamic programming algorithm that optimize subsequent measurements on the base of previous results.

As already pointed out in Section \ref{implementation}, an arbitrary projective measurement can be implemented with the Dolinar receiver setup. However, it is well known that the implementation of this receiver is challenging due to the requirements of a fast feedback from the photon counter to the local oscillator during the time slot, and suboptimal discrimination schemes results to be more practical in experimental setup. 

The dynamic programming algorithm can be set up with any local measurement scheme. In this section we consider the definition of an adaptive receiver scheme that uses a Generalized Kennedy receiver \cite{Takeoka2008} in each slot for the local binary discrimination, but leveraging on the dynamic programming to optimize the sequence of measurements. The resulting algorithm can again be represented with a tree as in Figure \ref{fig:binarytree}, but with the parameter $\beta_k=\angDP_k(\barz_k)$ instead of $\ang_k$, representing the time constant displacement to apply in the slot.

The transition probabilities of a Generalized Kennedy scheme are reported in equations \eqref{transitionGuha} with $\beta = \beta_k$ and $G=1$. In the framework setup, this transition probabilities are obtained from a pair of POVM, whose parameters can be obtained by inversion of \eqref{localTransPovm}.

Since the two expressions are not symmetric, we may consider two different associations between the outcomes $\{ \o,\ \l \}$ and the qubits $\{ \ket{\g_0}, \ \ket{\g_1} \}$. The association 
\beq
\left \{
\begin{aligned}
\p_{\o|0} & = \q_{\o|0} \\
\p_{\l|1} & = \q_{\l|\alpha}
\end{aligned}
\right.
\label{association1}
\eeq
comes directly from the definition of the Generalized Kennedy receiver. The other association, 
\beq
\left \{
\begin{aligned}
\p_{\o|0} & = \q_{\l|\alpha} \\
\p_{\l|1} & = \q_{\o|0}
\end{aligned}
\right.
\label{association2}
\eeq
can be obtained placing a displacement operation $D(-\alpha)$ before the receiver, therefore inverting the role of the ground state and the non-nulled coherent state.

Replacing the local measurements leads to minor changes in the dynamic programming algorithm. The reward-to-go function $\fTable_{\M-1}^\ast$ and the optimized parameter $\angDP_{\M-1}^\ast$ for the last measurement are evaluated as
\begin{align}
\fTable_{\M-1}^\ast(\s_{\M-1}) & = \operatorname*{max}_{\beta_{\M-1}} \ \fTable_{\M-1}(\s_{\M-1},\beta_{\M-1}), \label{initialfTableGK} \\
\angDP_{\M-1}^\ast(\s_{\M-1}) & = \operatorname*{argmax}_{\beta_{\M-1}} \ \fTable_{\M-1}(\s_{\M-1},\beta_{\M-1}), \label{initialPhiGK}
\end{align}
with
\begin{align}
& \fTable_{\M-1}(\s_{\M-1},\beta_{\M-1}) = \p_{\M} + \left[(\p_{\mx}+\p_{\M}) \sinh \alpha \beta_{\M-1} \right.\notag \\
& \qquad + \left. (\p_{\mx}-\p_{\M}) \cosh \alpha \beta_{\M-1} \right] \e^{-\beta_{\M-1}^2-\frac{|\alpha|^2}{4}}.
\end{align}
The transition probabilities $\{ \pProj_{\o|0},\ \pProj_{\l|1}\}$ of the projective measurements need to be replaced with  $\{ \p_{\o|0},\ \p_{\l|1}\}$. Both the possible associations \eqref{association1} and \eqref{association2} must be consider in the update equations \eqref{AC}-\eqref{BD}, as well as in the optimization of the first measurement in equations \eqref{ang1} and \eqref{fTable0}.

This adaptive receiver improves the performance of the Improved Conditional Nulling type I scheme due to the more generality obtained by the dynamic programming. On the other hand, the performance will be suboptimal since a suboptimal scheme is used for the local measurements.

\section{\label{results}Results and Numerical Issues}

In the previous Section we described the algorithm to optimize the sequences of parameters $\ang_k$ used by the adaptive receiver. We run the optimization algorithm for different cardinalities $\M$ of the PPM and for different values of the inner product $\inner$. For a fair comparison with respect to the other existing schemes, we compare the performances of the receiver architectures on the base of mean photon number in the coherent state $\ket{\alpha}$, i.e. $|\alpha|^2$, obtained by inversion of \eqref{def:inner}.

\begin{figure}
\centering
\includegraphics{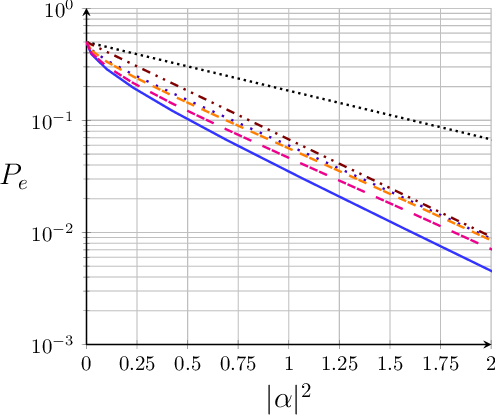}
\caption{(Color online) Performances comparison of different receiver schemes, for 2-PPM. The curves, from top to bottom, correspond to:  
\protect\tikz \protect\draw[line width=1pt,no markers,dotted] (0,-0.5ex) (0ex,0ex) -- (3.8ex,0ex); classical direct detection (black), 
\protect\tikz \protect\draw[line width=1pt,draw=red!50!black,no markers,dash pattern= on 4pt off 3pt on 1pt off 3pt on 1pt off 3pt] (0,-0.5ex) (0ex,0ex) -- (5ex,0ex); conditional nulling receiver (red), 
\protect\tikz \protect\draw[line width=1pt,draw=violet!50!blue,no markers,loosely dotted] (0,-0.5ex) (0ex,0ex) -- (3.8ex,0ex); type I improved conditional nulling scheme (violet), 
\protect\tikz \protect\draw[line width=1pt,orange,no markers,dash pattern= on 4pt off 1pt on 4pt off 4 pt] (0,-0.5ex) (0ex,0ex) -- (3ex,0ex);~type II improved conditional nulling scheme (orange),
\protect\tikz \protect\draw[line width=1pt,magenta,no markers,dash pattern= on 4pt off 1pt on 4pt off 1pt on 4pt off 6 pt](0,-0.5ex) (0ex,0ex) -- (4ex,0ex);~adaptive scheme with local Generalized Kennedy measurements (magenta),
\protect\tikz \protect\draw[line width=1pt,draw=blue!80!white,no markers,solid] (0,-0.5ex) (0ex,0ex) -- (4ex,0ex); quantum theoretical limit and adaptive with projective measurements (overlapped, blue). }
\label{fig:error_M2}
\end{figure}

A first result is that in the case of $\M=2$ the adaptive receiver with local projective measurements reaches the theoretical quantum limit. This is not surprising, because as already pointed out in \cite{Acin2005} in the case of binary discrimination of pure states an optimized sequence of local measurements suffices to implement the POVM for the optimal discrimination. In Figure \ref{fig:error_M2}  the performances of the classical receiver, conditional nulling, type I and type II schemes are compared with respect to the theoretical limit. The performance of the adaptive receivers surpass these schemes, and in the case of the adaptive receiver with projective measurements the performance overlaps with the theoretical one. 

\begin{figure}
\centering
\includegraphics{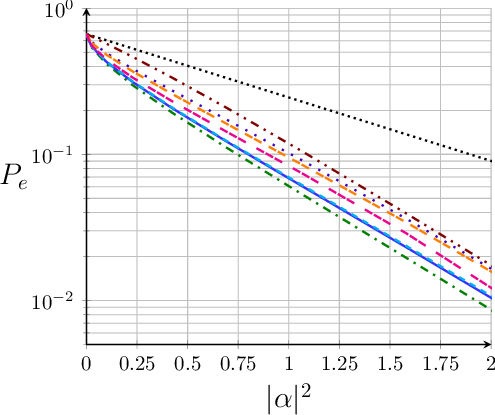}
\caption{(Color online) Performances comparison of different receiver schemes, for 3-PPM. The curves, from top to bottom, correspond respectively to: 
\protect\tikz \protect\draw[line width=1pt,no markers,dotted] (0,-0.5ex) (0ex,0ex) -- (3.8ex,0ex); classical direct detection (black), 
\protect\tikz \protect\draw[line width=1pt,draw=red!50!black,no markers,dash pattern= on 4pt off 3pt on 1pt off 3pt on 1pt off 3pt] (0,-0.5ex) (0ex,0ex) -- (5ex,0ex); conditional nulling receiver (red), 
\protect\tikz \protect\draw[line width=1pt,draw=violet!50!blue,no markers,loosely dotted] (0,-0.5ex) (0ex,0ex) -- (3.8ex,0ex); type I improved conditional nulling scheme (violet), 
\protect\tikz \protect\draw[line width=1pt,orange,no markers,dash pattern= on 4pt off 1pt on 4pt off 4 pt] (0,-0.5ex) (0ex,0ex) -- (3ex,0ex);~type II improved conditional nulling scheme (orange),
\protect\tikz \protect\draw[line width=1pt,magenta,no markers,dash pattern= on 4pt off 1pt on 4pt off 1pt on 4pt off 6 pt](0,-0.5ex) (0ex,0ex) -- (4ex,0ex);~adaptive scheme with local Generalized Kennedy measurements (magenta),
\protect\tikz \protect\draw[line width=1pt,draw=cyan,no markers,dashed] (0,-0.5ex) (0ex,0ex) -- (4ex,0ex); retraced forward adaptive scheme with projective measurements (cyan),
\protect\tikz \protect\draw[line width=1pt,draw=blue!80!white,no markers,solid] (0,-0.5ex) (0ex,0ex) -- (4ex,0ex); adaptive with projective measurements (blue),
\protect\tikz \protect\draw[line width=1pt,draw=green!50!black,no markers,dash pattern= on 4pt off 3pt on 1pt off 3pt] (0,-0.5ex) (0ex,0ex) -- (4ex,0ex); quantum theoretical limit (green). }
\label{fig:error_M3}
\end{figure}

\begin{figure}
\centering
\includegraphics{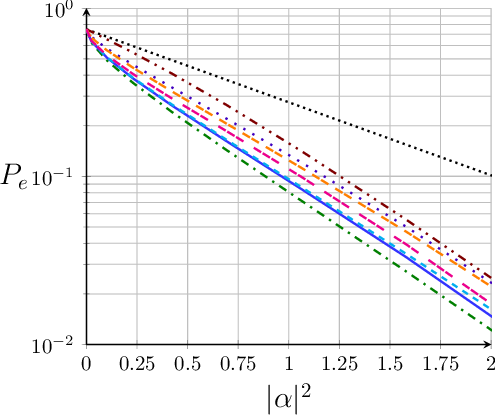}
\caption{(Color online) Performances comparison of different receiver schemes, for 4-PPM. The curves, from top to bottom, correspond respectively to: 
\protect\tikz \protect\draw[line width=1pt,no markers,dotted] (0,-0.5ex) (0ex,0ex) -- (3.8ex,0ex); classical direct detection (black), 
\protect\tikz \protect\draw[line width=1pt,draw=red!50!black,no markers,dash pattern= on 4pt off 3pt on 1pt off 3pt on 1pt off 3pt] (0,-0.5ex) (0ex,0ex) -- (5ex,0ex); conditional nulling receiver (red), 
\protect\tikz \protect\draw[line width=1pt,draw=violet!50!blue,no markers,loosely dotted] (0,-0.5ex) (0ex,0ex) -- (3.8ex,0ex); type I improved conditional nulling scheme (violet), 
\protect\tikz \protect\draw[line width=1pt,orange,no markers,dash pattern= on 4pt off 1pt on 4pt off 4 pt] (0,-0.5ex) (0ex,0ex) -- (3ex,0ex);~type II improved conditional nulling scheme (orange),
\protect\tikz \protect\draw[line width=1pt,magenta,no markers,dash pattern= on 4pt off 1pt on 4pt off 1pt on 4pt off 6 pt](0,-0.5ex) (0ex,0ex) -- (4ex,0ex);~adaptive scheme with local Generalized Kennedy measurements (magenta),
\protect\tikz \protect\draw[line width=1pt,draw=cyan,no markers,dashed] (0,-0.5ex) (0ex,0ex) -- (4ex,0ex); retraced forward adaptive scheme with projective measurements (cyan),
\protect\tikz \protect\draw[line width=1pt,draw=blue!80!white,no markers,solid] (0,-0.5ex) (0ex,0ex) -- (4ex,0ex); adaptive with projective measurements (blue),
\protect\tikz \protect\draw[line width=1pt,draw=green!50!black,no markers,dash pattern= on 4pt off 3pt on 1pt off 3pt] (0,-0.5ex) (0ex,0ex) -- (4ex,0ex); quantum theoretical limit (green). }
\label{fig:error_M4}
\end{figure}

\begin{figure}
\centering
\includegraphics{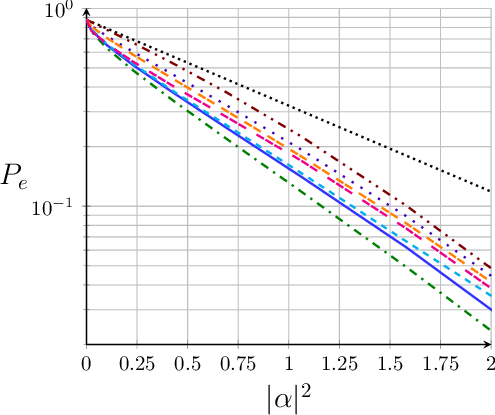}
\caption{(Color online) Performances comparison of different receiver schemes, for 8-PPM. The curves, from top to bottom, correspond respectively to: 
\protect\tikz \protect\draw[line width=1pt,no markers,dotted] (0,-0.5ex) (0ex,0ex) -- (3.8ex,0ex); classical direct detection (black), 
\protect\tikz \protect\draw[line width=1pt,draw=red!50!black,no markers,dash pattern= on 4pt off 3pt on 1pt off 3pt on 1pt off 3pt] (0,-0.5ex) (0ex,0ex) -- (5ex,0ex); conditional nulling receiver (red), 
\protect\tikz \protect\draw[line width=1pt,draw=violet!50!blue,no markers,loosely dotted] (0,-0.5ex) (0ex,0ex) -- (3.8ex,0ex); type I improved conditional nulling scheme (violet), 
\protect\tikz \protect\draw[line width=1pt,orange,no markers,dash pattern= on 4pt off 1pt on 4pt off 4 pt] (0,-0.5ex) (0ex,0ex) -- (3ex,0ex);~type II improved conditional nulling scheme (orange),
\protect\tikz \protect\draw[line width=1pt,magenta,no markers,dash pattern= on 4pt off 1pt on 4pt off 1pt on 4pt off 6 pt](0,-0.5ex) (0ex,0ex) -- (4ex,0ex);~adaptive scheme with local Generalized Kennedy measurements (magenta),
\protect\tikz \protect\draw[line width=1pt,draw=cyan,no markers,dashed] (0,-0.5ex) (0ex,0ex) -- (4ex,0ex); retraced forward adaptive scheme with projective measurements (cyan),
\protect\tikz \protect\draw[line width=1pt,draw=blue!80!white,no markers,solid] (0,-0.5ex) (0ex,0ex) -- (4ex,0ex); adaptive with projective measurements (blue),
\protect\tikz \protect\draw[line width=1pt,draw=green!50!black,no markers,dash pattern= on 4pt off 3pt on 1pt off 3pt] (0,-0.5ex) (0ex,0ex) -- (4ex,0ex); quantum theoretical limit (green). }
\label{fig:error_M8}
\end{figure}

As the cardinality $\M$ increases, the performance of the adaptive receiver with local projective measurements slightly moves away from the theoretical optimum. In Figure \ref{fig:error_M3}, \ref{fig:error_M4} and \ref{fig:error_M8} the performance of the existing and adaptive receivers are compared for $\M=3$, $\M=4$ and $\M=8$ respectively. The trend is the same in all the figures: the adaptive schemes outperform the existing conditional nulling, type~I and type~II receiver, placing the error probability curves between these and the theoretical limit. The adaptive schemes maintain the gap with respect to type~I and type~II even around $|\alpha|^2=2$, where these schemes get close to the conditional nulling performances. In addition, the performance of our scheme gets really close to the theoretical limit for low mean photon number.

The evaluation of the dynamic programming algorithm can be really demanding, in particular the evaluation of $\fTable_k^\ast$ for all possible system states $\s_{k}$ may require a non trivial amount of computational time and memory. In addition, this evaluation must be repeated from $k~=~\M~-~1$ down to $k=1$. 

Since a numerical procedure is required to evaluate $\fTable_k^\ast$ at each step, the set $\{(u,v), 0\leq u \leq 1, 0\leq v \leq 1, u+v\leq 1\}$ of the system state space $\S_k$ needs to be discretized in a two dimensional grid. As a consequence, the search of the optimal parameters $\angDP_k^\ast$ in \eqref{angK} and the evaluation of $\fTable_{k}^\ast$ in \eqref{updatefTable} requires to approximate the system state $\s_{k+1}$ in the grid when considering $\fTable_{k+1}^\ast(\s_{k+1})$.
The issue of this approximation spread out in successive evaluations of $\fTable_k^\ast$, especially in the case of poorly discretized grid, where we encounter bad (even unfeasible) results for high values of $\M$ and $|\alpha^2|$. In our optimization, we use a discretization with at least a grid step of $10^{-3}$ for each side of the unit square that includes the set of $(u,v)$ in \eqref{def:state}.

Some considerations can be done in order to lighten the computation. The first consideration is that for different cardinality $\M$, the sequence of tables $\fTable_k^\ast$ to be calculated are the same. This means that if we want to evaluate the performances of the adaptive receiver for a PPM with cardinality $\M_1 < \M_2 < \ldots < \tilde{\M}$, we can calculate the table sequence $\fTable_{\tilde{\M}}^\ast, \fTable_{\tilde{M}-1}^\ast, \ldots, \fTable_1^\ast$ for the maximum cardinality $\tilde{\M}$. In evaluating the performances of the other modulation cardinality $\M_i$, we only need to start from a different initial system state, that is 
\beq
\s_0 = \left(\frac{1}{\M_i},\frac{1}{\M_i} , \cdot \right)
\eeq
end evaluate the probability of correct decision as
\begin{align}
\Pc = \operatorname*{\max}_\ang & \quad \fTable_{\tilde{M}-\M_i+1}^\ast\left(\frac{\pProj_{\o|1}(\ang)}{\M_i}, \frac{\pProj_{\o|0}(\ang)}{\M_i}, 1 \right)  \notag \\
& \qquad +\fTable_{\tilde{M}-\M_i+1}^\ast\left(\frac{\pProj_{\l|1}(\ang)}{\M_i}, \frac{\pProj_{\l|0}(\ang)}{\M_i}, 1 \right) 
\end{align}

In addition, as already pointed out and proved in Lemma \ref{probabilityM}, before the $k+1$-th measurement the joint probability of the $\M-k$ symbols $\x=k+1, k+2, \ldots, \M$ are the same. This means that considering the variables $\p_\mx, \p_\M$ of the system state $\s_{k}$ that define the entries of $\fTable_{k}^\ast$, it results $\p_\M\leq\frac{1}{\M-k+1}$, therefore reducing the elements of the set $\{(u,v)\}$ of $\S_k$ to evaluate in 
\beq
\left  \{(u,v), u+v\leq 1, 0\leq u \leq1, 0\leq v \leq \frac{1}{\M-k+1} \right \}. 
\eeq

Furthermore, if we are interested in the performance for a single value of the cardinality $\M$, we can take advantage of Lemma \ref{decreasing} in Appendix \ref{preliminar}. Since the elements $\p_\mx$ and $\p_\M$ are \emph{joint} probabilities of symbols with the outcome sequence, and since they start from the value $1/\M$, we can restrict the grid to evaluate for table $\fTable_k^\ast$ to consider only the set
\beq
\left \{(u,v), 0\leq u \leq \frac{1}{\M}, 0\leq v \leq \frac{1}{\M} \right \}. 
\eeq

In order to understand the consequence of the approximation of the system state space $\S_k$, we check the performances of the dynamic programming retracing all the parameters path for each measurement, evaluating the final joint probabilities and calculating the probability of correct decision as the sum in \eqref{estimationStage}. Due to the discretization of $\S_k$ as a grid, the performance obtained retracing the parameters tree can be slightly different with respect to the performances of dynamic programming. The performances of this \emph{forward} path retracing are depicted in Figures \ref{fig:error_M2}, \ref{fig:error_M3}, \ref{fig:error_M4} and \ref{fig:error_M8} in a dashed (cyan) line. As you can see, for lower cardinality it coincides with the prediction, but the gap spreads out as $\M$ increases, especially for high $|\alpha|^2$ (see for example Figure \ref{fig:error_M4}). 

In Figure \ref{fig:error_M8}, we managed to keep the performances of the forward retracing close to the predicted one for $\M=8$  by discretizing the grid $[0,1/8] \times [0,1/8]$ with $10^3 \times 10^3$ samples.

\section{\label{conclusions}Conclusion}

In the present work we have studied the design of quantum receivers for Pulse Position Modulation. 

By the PPM signal structure, we could describe the overall transmitted quantum states in the symbol time interval as sequences of quantum states  in shorter temporal slots in a tensorial product. The signal measurement is then reformulated as a sequence of shorter measurements, one in each slot, that allows to design adaptive receiver scheme. 

We move to an isomorphic representation of the quantum states in terms of qubits. The description of the existing receiver architecture in this framework highlights the limitations in terms of outcomes probabilities. We propose a more general adaptive receiver structure, where the measurement in each slot is a function of all the previous outcomes and the time evolving joint probabilities of the symbols with the outcomes sequence. 

We propose an optimization of such adaptive scheme by means of dynamic programming, providing a description of the algorithm to evaluate the performance of the adaptive receiver and to calculate the measurement in each slot. We consider both (optimal) projective and (suboptimal) Generalized Kennedy local measurements in each slot. The probability of error, although it does not reach the theoretical quantum limit except for $\M>2$, significantly outperforms the existing receiver schemes. 

As a concluding remark, adaptive receiver seem to be the way to follow to achieve better performances for communication purpose, thanks to the possibility to embed the information of previous outcomes and improve subsequent measurements.

This work has been carried out within the Strategic-Research-Project QUINTET of the Department of Information Engineering, University of Padova and the Strategic-Research-Project QUANTUMFUTURE of the University of Padova.

\appendix

\section{\label{preliminar} Useful Lemmas}

In this Appendix we
give some useful Lemmas that helps to better understand the optimization algorithm.

\begin{Lemma}
\label{discrimination2}
Let $\p_0,\p_1, \ \p_0+\p_1\leq1$ be the a priori probability of two symbols $i=0,1$ associated to the quantum states $\ket{\g_0}, \ket{\g_1}$ respectively. The maximum probability of correct discrimination between $i=0$ and $i=1$ achievable with measurement operators \eqref{operatorH} defined by $\ang$ is 
\beq
\frac{1}{2} \left[ \p_0 + \p_1 + \sqrt{(\p_0 + \p_1)^2-4 \p_0 \p_1 \inner}\right],
\label{PcDiscrimination2}
\eeq
and the optimal angle is defined by
\beq
\ang^\ast = \frac{1}{2} \phase{\cos 2 \theta(\p_0 - \p_1)+ \imaginary \sin 2 \theta (\p_0 + \p_1)}.
\label{optAngle}
\eeq
\end{Lemma}
\proof Consider the measurement operators \eqref{operatorH}, and without loss of generality associate the outcomes $\z=\o$ with the estimation of $i=0$ and $\z=\l$ with $i=1$.
The probability of correct discrimination can be written as
\begin{align}
\Pc & 
= \Prob{\z=\o|i=0} \p_0 + \Prob{\z=\l|i=1} \p_1 \notag \\
& = \cos^2 (\theta -\ang) \p_0 + \sin^2(\theta + \ang) \p_1 \notag \\
& = \frac{1+\cos(2 \theta - 2\ang)}{2} \p_0 + \frac{1-\cos(2\theta + 2 \ang)}{2} \p_1 \label{eqLemma1}
\end{align}
with transition probabilities defined by \eqref{localTrans}. The maximization of \eqref{eqLemma1} with respect to the angle $\ang$ leads to the relation
\beq
\tan 2 \ang = \tan 2 \theta \frac{\p_0 + \p_1}{\p_0 - \p_1},
\eeq
solved by the angles verifing 
\beq
\begin{array}{rcl}
\sin 2 \ang & = & \pm \frac{\displaystyle \p_0 + \p_1}{\displaystyle  \sqrt{R}} \sin 2 \theta ,\\[10pt]
\cos 2 \ang & = & \pm \frac{\displaystyle  \p_0 - \p_1}{\displaystyle  \sqrt{R}} \cos 2 \theta,
\end{array}
\label{Angle}
\eeq
with $R$ a normalization term, 
\beq
R = \cos^2 2 \theta (\p_0-\p_1)^2 + \sin^2 2\theta (\p_0 + \p_1)^2 .
\eeq
Expression \eqref{Angle} with the plus sign corresponds to the point of maximum, and the thesis \eqref{optAngle} follows.
Substituting \eqref{optAngle} in \eqref{eqLemma1} gives
\begin{align}
\Pc & = \frac{1}{2} \left[ \p_0 +\p_1 + \sqrt{R}\right] \notag \\
& = \frac{1}{2} \left[ \p_0 +\p_1+  \sqrt{(\p_0 + \p_1)^2 -4 \p_0 \p_1 \cos^2 2 \theta }\right] .
\end{align}
\qed

\begin{cor}
\label{discriminationCor}
If the quantum states associated to the two symbols $i=0,1$ is the same, i.e. $\ket{\g_0} = \ket{\g_1}$, the maximal probability of correct discrimination between $i=0$ and $i=1$ is the maximum of their a priori probability, 
\beq
\Pc = \max \{\p_0, \p_1\}.
\label{PcDiscriminationMax}
\eeq
\end{cor}
The Corollary follows from $\inner = |\inn{\g_0}{\g_1}|^2 =1$. Intuitively, this means that in the case of the same quantum states, we cannot discriminate the symbols $i=0, 1$ better that their a priori distribution. 

Moreover, since \eqref{PcDiscrimination2} is always non lower than \eqref{PcDiscriminationMax}, at the last measurement it is always better to discriminate between the last symbol $\x=\M$ and a previous one $\x=i<\M$. This is reasonable since the last slot would eventually deliver information about the last symbol, and it is useless to discriminate previous ones. 

In addition, expression \eqref{PcDiscrimination2} is monotonically increasing with the probabilities $\p_0, \p_1$, such that it is always better to compare the symbol $\x=i<\M$ with highest a priori probability.

\begin{Lemma}
\label{orderingM}
The relative ordering of the a priori probabilities of symbols $i,j \neq k$ before the $k$-th measurement is maintained in the a posteriori distribution, independently of the outcome $\z_k$.
\end{Lemma}
\proof Consider two symbols $\x=i$ and $\x=j$, $i,j \neq k$ with a priori probabilities $\p_{i,\z_{k-1}},\p_{j,\z_{k-1}}$ before the $k$-th measurement, with $\p_{i,\z_{k-1}}>\p_{j,\z_{k-1}}$. Both symbols $i,j$ has a quantum state $\ket{\g_0}$ in the $k$-th position of $\ket{\barg_i},\ \ket{\barg_j}$. Hence, the transition probability is the same, 
\beq
\Prob{\z_k|i} = |\inn{\m_{\z_k}}{\g_0}|^2 =\Prob{\z_k|j}
\eeq
and the joint probabilities are multiplied by the same factor
\beq
\p_{i,\z_k} = |\inn{\m_{\z_k}}{\g_0}|^2 \p_{i,\z_{k-1}}, \quad \p_{j,\z_k} = |\inn{\m_{\z_k}}{\g_0}|^2 \p_{j,\z_{k-1}},
\eeq
and hence the a posteriori distribution of $\x=i,j\neq k$ reflects the same relative ordering of the priori. \qed

This Lemma is a consequence of the fact that $i,j\neq k$ has the same quantum states $\ket{\g_0}$ in position $k$. The $k$-th measurement does not give any information about the discrimination between symbols $i,j\neq k$ because they all behave in the same way with respect to a measurement in this slot. The $k$-th slot can give information only for the the discrimination of the $k$-th symbol, whether it is more likely or not respect the other.

\begin{cor}
\label{orderingK}
Given an outcome sequence $\barz_k$, the relative ordering of the joint probabilities of symbols $\x=1, \ldots, k$ is maintained in the joint probabilities after measurement $k+1, \ldots, \M$, independently of the outcomes $\z_{k+1}, \ldots, \z_\M$.
\end{cor}
\proof The quantum states of symbols $\x=1, \ldots, k$ in position $k+1, \ldots, \M$ of the tensor product are $\ket{\g_0}$, so the joint probabilities after measurement $l>k$ is
\begin{align}
& \p_{i, [\barz_k \z_{k+1} \ldots \z_l]} = \notag \\
& \quad \Prob{\z_l|\z_{l-1}, \ldots,\barz_k ,\ket{\g_0}} \cdot \ldots \cdot \Prob{\z_{k+1}|\barz_k ,\ket{\g_0}} \p_{i, \barz_k} 
\end{align}
that is, all the joint probabilities are multiplied by the same factors. \qed

\begin{Lemma}
\label{probabilityM}
The joint probabilities of symbols $\x=k+1, k+2, \ldots, \M$ with $\barz_l$ after any measurement in the $l$-th slot,  $l = 1, \ldots, k$, verify $\p_{k+1,\barz_l} = \p_{k+2,\barz_l} = \ldots = \p_{\M,\barz_l}$.
\end{Lemma}
\proof Consider a measurement $l \in 1, \ldots, k$. The joint probabilities of symbols $\x=k+1, k+2,\ldots, M$ with the outcomes vector $\barz_l$ can be calculated as
\beq
\p_{\x,\barz_l}=\Prob{\z_l|\barz_{l-1},\x}\Prob{\z_{l-1}|\barz_{l-2},\x} \cdots \Prob{\z_1|\x} \frac{1}{M},
\eeq
but since each symbol $\x=k+1,\ldots,M$ has a quantum state $\ket{\g_0}$ in position $1,\ldots,l,\ l\leq k$, the conditional probabilities give
\beq
\p_{\x,\barz_l}=|\inn{\m_{\barz_l}}{\g_0}|^2|\inn{\m_{\barz_{l-1}}}{\g_0}|^2 \cdots |\inn{\m_{\barz_1}}{\g_0}|^2\frac{1}{M},
\eeq
and hence $\p_{k+1,\barz_l} = \p_{k+2,\barz_l} = \ldots = \p_{M,\barz_l}$. \qed

Intuitively, since all symbols $\x=k+1, k+2,\ldots, M$ have the quantum state $\ket{\g_0}$ in position $1, \ldots, k$, we cannot gain any information about their discrimination on the base of the first $k$ outcomes, and therefore their joint probabilities are the same.


\begin{Lemma}
Joint probabilities are non increasing in subsequent measurements, and always lower than the a priori probability $\frac{1}{M}$.
\label{decreasing}
\end{Lemma}
\proof Writing the joint probability with the conditional chain rule
\beq
\p_{i,\barz_k} = \Prob{\z_k|\barz_{k-1},i} \Prob{\z_{k-1}| \barz_{k-2},i} \cdots \Prob{\z_1|i} \frac{1}{M} 
\eeq
we see that after each measurement, the joint probabilities are updated with the transition probabilities depending upon the outcome. Since the transition probabilities are not greater that 1, they are non increasing, and it is clear that $\p_{i,\barz_k}\leq \frac{1}{M}$. \qed


\bibliographystyle{apsrev}
\bibliography{ppmBiblio}

\end{document}